\documentclass[onecolumn]{revtex4}
\usepackage{import}
\usepackage[utf8]{inputenc}
\usepackage{amsmath,amssymb,graphicx,bbm,mathrsfs,bm}
\usepackage[usenames, dvipsnames]{color}
\usepackage{float}
\usepackage{graphicx,setspace}
\usepackage{caption}

\usepackage{siunitx}
\restylefloat{table}
\begin{document}
\setstretch{1.8}
\title{ECCENTRIC: a fast and unrestrained approach for high-resolution in vivo metabolic imaging at ultra-high field MR}
\author{Antoine Klauser${}^{1,2,5,6,}$\footnote{Corresponding author, antoine.delattre-klauser@siemens-healthineers.com}, Bernhard Strasser${}^{1,3}$, Wolfgang Bogner${}^3$, Lukas Hingerl${}^3$, Sebastien Courvoisier${}^{5,6}$, Claudiu Schirda${}^4$, Francois Lazeyras${}^{5,6}$, Ovidiu C. Andronesi${}^{1,}$\footnote{Corresponding author, oandronesi@mgh.harvard.edu.} }

\affiliation{${}^1$Athinoula A. Martinos Center for Biomedical Imaging, Department
of Radiology, Massachusetts General Hospital, Harvard Medical School, Boston}
\affiliation{${}^2$Advanced Clinical Imaging Technology, Siemens Healthcare AG, Lausanne, Switzerland}
\affiliation{${}^3$High‐Field MR Center, Department of Biomedical Imaging and Image‐guided Therapy, Medical University of Vienna, Vienna, Austria}
\affiliation{${}^4$Department of Radiology, University of Pittsburgh School of Medicine,Pittsburgh, Pennsylvania, USA}
\affiliation{${}^5$Department of Radiology and Medical Informatics, University of Geneva, Switzerland}
\affiliation{${}^6$CIBM Center for Biomedical Imaging, Switzerland}

\date{\today}

\begin{abstract}
A novel method for fast and high-resolution metabolic imaging, called ECcentric Circle ENcoding TRajectorIes for Compressed sensing (ECCENTRIC), has been developed and implemented at 7 Tesla MRI. ECCENTRIC is a non-Cartesian spatial-spectral encoding method optimized to accelerate magnetic resonance spectroscopic imaging (MRSI) with high signal-to-noise at ultra-high field. The approach provides flexible and random ($k,t$) sampling without temporal interleaving to improve spatial response function and spectral quality. ECCENTRIC needs low gradient amplitudes and slew-rates that reduces electrical, mechanical and thermal stress of the scanner hardware, and is robust to timing imperfection and eddy-current delays. Combined with a model-based low-rank reconstruction, this approach enables simultaneous imaging of up to 14 metabolites over the whole-brain at 2-3mm isotropic resolution in 4-10 minutes. In healthy volunteers ECCENTRIC demonstrated unprecedented spatial mapping of fine structural details of human brain neurochemistry. This innovative tool introduces a novel approach to neuroscience, providing new insights into the exploration of brain activity and physiology.

\end{abstract}

\maketitle


\section{Introduction}
Magnetic Resonance Spectroscopic Imaging (MRSI) is a well-established molecular MR imaging modality, facilitating non-invasive exploration of \textit{in vivo} metabolism in both human and animal models without the use of ionizing radiation. In particular, ${}^1$H-MRSI can simultaneously image up to 20 brain metabolites, providing quantification of steady-state concentrations \cite{Maudsley2020} and the dynamic change of concentrations under functional tasks \cite{Mullins2018,Bednarik2021}. In addition to measuring intrinsic metabolism without the need of contrast agents, MRSI can probe metabolic enzymatic rates that are not accessible by nuclear imaging techniques such as PET and SPECT \cite{Davis2020}. Many studies demonstrated significant value of MRSI for neuroscience \cite{Oz2014}, but the performance of current MRSI is severely lacking behind other MRI methods, which limits its use and wider adoption.

Among MRI modalities, MRSI is poised to benefit the most from ultra-high field (UHF $\geq$ 7T) due to increased spectral dispersion and signal-to-noise ratio (SNR). MRSI using very short echo-time ($\approx$ 1ms) free induction decay (FID) excitation \cite{Henning2009,Boer2012,Bogner2012} has large potential for metabolite imaging due to its high SNR. Nevertheless, MRSI is hampered by significant limitations, including low resolution and long scan times required for the acquisition of the 4D($k,t$) spatial-temporal space \cite{Bogner2021}. This highlights a pressing demand for acceleration strategies in high-resolution MRSI to overcome these challenges. This is especially pertinent in the context of high-resolution whole-brain MRSI, where conventional phase-encoding acquisition schemes would necessitate several hours. Acceleration of UHF MRSI has been shown by parallel imaging such as SENSE, GRAPPA and CAIPIRINHA with uniform undersampling \cite{Strasser2017,Hangel2017,Nassirpour2018b}, or by Compressed Sensing (CS) with random undersampling \cite{Nassirpour2018a}, but these techniques generally don’t allow acceleration factors (AF) above 6-10 for MRSI. Additionally, spatial-spectral encoding (SSE) techniques introduce additional prospects for accelerating UHF MRSI. By combining spatial-spectral encoding with undersampling higher accelerations ($AF>50$) of UHF MRSI may be achieved \cite{Moser2019a,Saucedo2021,Ma2016}.

So far, SSE has been demonstrated at ultra-high fields using either Cartesian (echo-planar) \cite{An2018,Nam2022,Weng2022} and non-Cartesian (spirals, rosettes, concentric circles) \cite{Chiew2018,Hingerl2019,Moser2019a,Esmaeili2021} $k$-space trajectories. However, the methods shown to date have an important limitation that requires the use of temporal interleaves to achieve broad spectral bandwidth and high-spatial resolution. As the field strength increases, fulfilling these requirements becomes even more challenging. Temporal interleaving prolongs the acquisition time and creates spectral side-bands that degrade SNR and overlap with metabolite spectra \cite{Bogner2021}. 

Here we addressed limitations of the current MRSI methods to significantly improve metabolic imaging at ultra-high field with higher acceleration, spatial resolution, SNR and spectral quality. The main design specification for our method was to push the spatial-temporal limits of metabolic imaging for advanced neuroscience applications. In addition, we optimized robust and reproducible performance for quantitative science.
To accomplish these aims, we developed ECCENTRIC method (ECcentric Circle ENcoding TRajectorIes for Compressed sensing) that benefits from: 1) improved pseudo-random sampling for compressed sensing with non-Cartesian trajectories, 2) flexible sampling of the 4D ($k,t$) space for optimal SNR, 3) reduced demand on the gradient system for spectral quality, and 4) low-rank reconstruction for dimensionality reduction and denoising of data.

The performance of the new acquisition-reconstruction scheme was first investigated by simulations and in structural-metabolic phantoms, and subsequently evaluated $\textit{in vivo}$ in healthy subjects.

\section{Theory}

Circular trajectories, including rosettes and concentric circles, provide several advantages over spiral and echo-planar trajectories in MRSI and MRI \cite{Posse1994,Adalsteinsson1998,Furuyama2012,Schirda2018}. By design, ECCENTRIC's circular trajectories need smaller diameter than rosettes and concentric circles, which help achieve high-resolution and large spectral bandwidth at ultra-high field without temporal interleaving. ECCENTRIC trajectories are produced by readout gradient wave-forms that: 1) do not need rewinding which eliminates dead-time and the associated loss in SNR per unit time, 2) permit high matrix sizes with limited gradient amplitude, 3) have constant and moderate gradient slew-rate that is not demanding for patients nor gradient hardware, hence minimizing nerve stimulation and artifacts caused by eddy currents, gradient warming, field drift and mechanical resonances.

\begin{figure}[t]
\begin{center}
\includegraphics[width=1\textwidth]{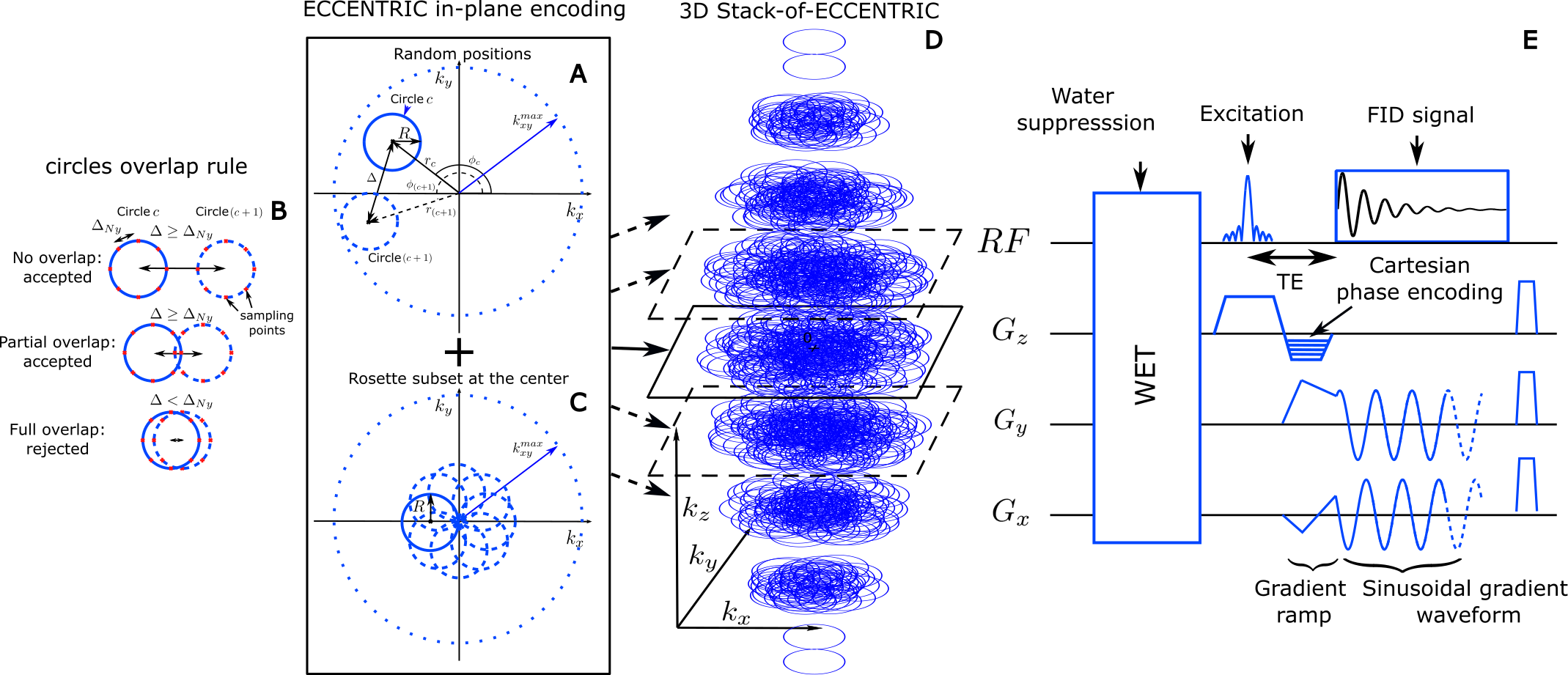}
\caption{ 3D ECCENTRIC sampling and acquisition. (A) circle center positions are parameterized in polar coordinates $(r_c,\phi_c)$ that are chosen randomly in the ranges $r_c\in [0,\max(k_{xy}^{max}-r,r)]$ and $\phi_c\in [0,2\pi]$. Two consecutive circles ($c$ and $c+1$) must respect the overlap rule described in (B): the distance between their respective centers, $\Delta$, must be greater or equal to the Nyquist distance, $\Delta_{Ny}$. (C), to satisfy a systematic full sampling of the $k$-space center, a small subset ($<5\%$) of circles is positioned in rosette pattern in each ECCENTRIC encoding planes. (D), 3D $k$-space sampling is achieved by a stack of ECCENTRIC encoding planes with variable $k_{x,y}^{max}$ to realize an ellipsoid coverage. (E) Diagram of the 3D FID-ECCENTRIC sequence. First, a 4-pulse WET water suppression technique is used, followed by the Shinnar–Le Roux optimized excitation pulse. After the excitation, the Cartesian encoding is performed along the z-axis, simultaneously to the gradient ramp along the x- and y-axes to reach the desired $k$-space off-center position and velocity. Finally, a sinusoidal gradient wave-form is applied along the x- and y-axis during acquisition to produce the circular trajectory. 
 }
\label{fig:Sampling}
\end{center}
\end{figure}

Moreover, the implementation of CS acceleration \cite{Donoho2006,Candes2006} relies on two prerequisites. The first is that the signal or image exhibits sparsity in a known transform domain \cite{Lustig2007,Knoll2011}. The second is that the data are randomly undersampled, which can be achieved by random undersampling of the $k$-space in MRI applications. 
To enable the random sparse undersampling necessary for CS, we utilized a novel approach where successive circular trajectories are randomly positioned in $k$-space, rather than using regular patterns such as rosette, concentric, or uniformly distributed circles. The acquisition strategy of ECCENTRIC is illustrated in Fig.\ref{fig:Sampling}. The circle centers' polar coordinates $(r_c, \phi_c)$ are chosen randomly with a uniform probability within the ranges $r_c\in [0,\max(k_{x,y}^{max}-R,R)]$ and $\phi_c\in [0,2\pi[$ (Fig. \ref{fig:Sampling}{\it A}).  Here, $R$ represents the circle radius, $k_{x,y}^{max}$ is the largest in-plane $k$-space coordinate (assuming the same spatial resolution along all axial plane directions). The majority of circles are placed randomly as shown with two successive circles ($c$ and $c+1$) in the sketch Fig. \ref{fig:Sampling}{\it A}, but with the constraint to avoid significant overlap between circles (Fig.\ref{fig:Sampling} {\it B}): the distance between the centers of each circle, $\Delta$, must be larger or equal to the Nyquist distance (the inverse of the FoV size). When $\Delta : < 2R$, there is partial overlap between circles; however, this redundant sampling is predominantly concentrated in the central region of $k$-space, enabling an enhancement in SNR. In addition to the random pattern, a small subset of circles ($<5\%$ of the total number) positioned in rosette fashion is acquired in the center of $k$-space (Fig.\ref{fig:Sampling} {\it C}). This ensures complete sampling of the center of $k$-space, which is beneficial for SNR and reconstruction performance \cite{Knoll2011,Lustig2007} with negligible effect on acquisition time. The homogeneous random distribution of circle polar coordinates results intrinsically in a pseudo-random $k$-space sampling with density following $1/\|k\|$ outside of the rosette sampled central region.

The ECCENTRIC design with free choice of circle's radius and center in the range of $0$ to $\frac{k_{x,y}^{max}}{2}$ provides complete flexibility in selecting the matrix size, field-of-view (FoV), and spectral bandwidth. This design design allows to not exceed the gradient hardware's slew-rate and amplitude limitations over a wide parameter space. ECECNTRIC combined with compressed-sensing  fulfills better the random undersampling requirement compared to echo-planar \cite{Hu2008,Otazo2009,Hu2010,Kampf2010,Iqbal2016,Santos-Diaz2019}, spiral \cite{Chatnuntawech2015} and radial \cite{Saucedo2021} trajectories.

To extend ECCENTRIC to 3D $k$-space sampling, a stack of ECCENTRIC is employed with circles randomly placed in the $k_x$-$k_y$ planes, while $k_z$ is encoded using Cartesian phase-encoding (Fig.\ref{fig:Sampling} {\it D}). The 3D $k$-space can be covered using spherical or ellipsoid coverage, where the in-plane $k$-space boundary is defined as $k_{x,y}^{max} = \frac{n }{2\: FoV}\sqrt{1-(k_z/k_z^{max})^2}$, with $FoV$ representing the FoV size and $n $ the spatial resolution.
To achieve complete sampling in a single $k_x$-$k_y$ plane, the number of needed ECCENTRIC circles can be derived similar to rosette encoding that requires $\frac{\pi n}{ 2}$ circles \cite{Schirda2009}. To maintain the same number of sampling points, the total number of ECCENTRIC circles required for complete sampling of a single stack is $ \frac{\pi n k_{x,y}^{max} }{2 \: R}$. To achieve circle encoding with off-center position $(r_c,\phi_c)$, a brief gradient ramp is used to gain an initial momentum $(k_x,k_y)$ position and the necessary velocity. This process is done at the same time as the slab excitation rewinder overlapped by $z$-phase encoding and does not increase the echo time, as shown in Fig.\ref{fig:Sampling}{\it E}.
In implementing CS acceleration,  the total number of ECCENTRIC circles $N_c$ is reduced uniformly across the stacks by a factor of AF.

\begin{figure}[!th]
\centering
\includegraphics[width=\textwidth]{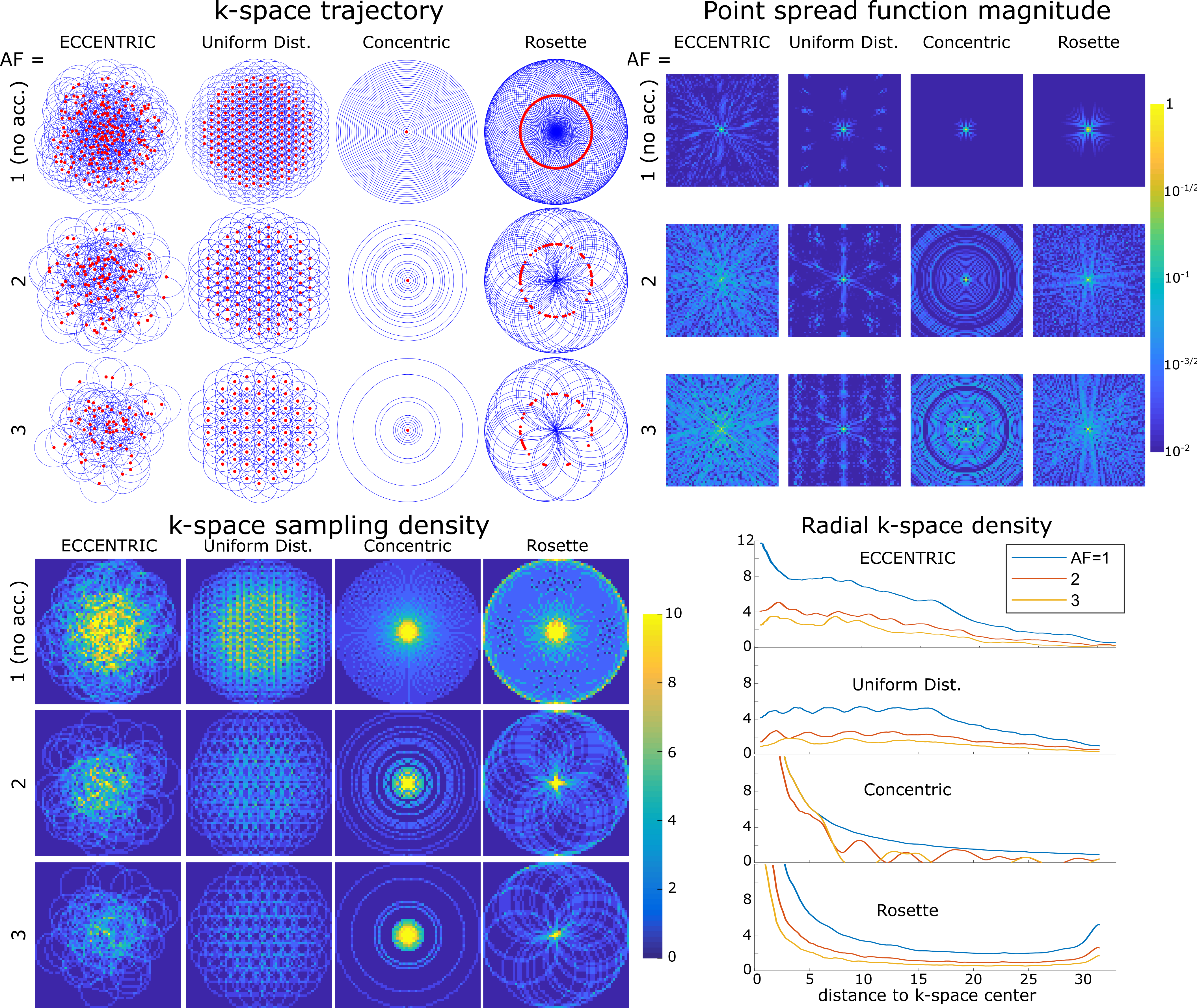}
\caption{Comparative analysis of circular $k$-space encodings. Top left,$k$-space trajectories for ECCENTRIC, uniform distributed circles, concentric circles and rosette trajectories for a $64\times64$ encoding matrix. Red dots indicate the circle center positions. The acceleration factors AF=$1,2,3$ correspond to ECCENTRIC and uniform distributed circles trajectories with $202$, $101$ and $67$ circles respectively; $31$, $16$ and $11$ concentric circles; $101$, $51$ and $34$ Rosette circles. Top right, the point spread function (PSF) calculated for each trajectory and acceleration on a log-scale highlight the presence of incoherent and coherent aliasing patterns. Bottom, the sampling density for the same trajectories and AFs, represented in the 2D $k$-space (left) and along a radial projection (right).}

\label{fig:KSpace}
\end{figure}

In Fig.\ref{fig:KSpace} a comparison is made between ECCENTRIC, uniform distributed circles trajectory,
 concentric circles and rosette sampling. The trajectory and sampling density in the $k$-space for each pattern and acceleration factor highlight the differences in sampling distribution. The density distribution of ECCENTRIC falls between the density profile of rosette/concentric circles, characterized by a high-density singularity at the $k$-space center, and the uniformly distributed circles, which exhibit a flatter profile extending to the periphery of the $k$-space. While rosette and concentric circle trajectories provide a favorable sampling density at the center of $k$-space, they require temporal interleaving that increase the acquisition time and create sidebands and spectral artefacts which are detrimental to the metabolite signal. Uniform distributed circles have less desirable $k$-space density, but does not need temporal interleaving similar to ECCENTRIC. Thus, ECCENTRIC offers a favorable compromise by achieving high density in the center of $k$-space for improved SNR while avoiding the need for temporal interleaving.

The point spread function (PSF) simulations in Fig.\ref{fig:KSpace} indicate that ECCENTRIC is inherently suited for random undersampling required for good CS performance \cite{Lustig2007}. The PSF reflects the interference between voxels in image space resulting from undersampling. A PSF that shows pseudo-random incoherent pattern is more suited for CS acceleration. Simulations reveal an incoherent pattern for PSF of ECCENTRIC due to the pseudo random $k$-space sampling, and the PSF pattern spreads with the increasing acceleration factor but conserve the pseudo-random behavior with undersampling (Fig. \ref{fig:KSpace}). In comparison, uniform distributed circles, concentric circles, and rosette trajectories have more compact PSFs for fully sampled acquisitions and their PSFs exhibit coherent patterns when undersampling is applied, which is less favorable for CS acceleration. The PSF was computed on a $64\times64$ matrix and obtained from a single point source \cite{Mareci1991} that was encoded in $k$-space and then reconstructed with Non-uniform Fourier transform with Voronoi's partition density compensation \cite{Rasche1999}. 

Due to the non-uniformity and sparsity of the sampling, a specific model is necessary to reconstruct 4D ($k,t$) data of ECCENTRIC into image-frequency space. In previous studies \cite{Klauser2019, Klauser2021,Klauser2022}, we demonstrated the effectiveness of CS-SENSE-LR model that combines partial-separability (or low-rank) with Total-Generalized-Variation (TGV) constraint for reconstructing Cartesian $k$-space data acquired with random undersampling, leading to improved SNR. Here, we extended the CS-SENSE-LR approach to incorporate non-uniform Fourier sampling necessary for reconstructing ECCENTRIC data.
Defining the discrete MRSI data in image space to be $\rho$ as an $N_{\mathbf{r}}$ by $T$ array (with $N_{\mathbf{r}}$ the number of spatial points and $T$ the number of sampling time points), the low-rank hypothesis on the magnetization assumes that the MRSI data can be separated into a small number of spatial and temporal components :
\begin{equation}\label{eq:LR}
 \rho = \mathbf{U}\mathbf{V}
\end{equation}
where $\mathbf{U}$ is a $N_{\mathbf{r}}$ by $K$ array and $\mathbf{V}$ a $K$ by $T$ array, with $K$ the rank of the low-rank model. These components are retrieved by CS-SENSE-LR reconstruction solving the inverse problem 
 \begin{eqnarray}\label{eq:Recon}
 \arg \min_{ \mathbf{U},\mathbf{V},\mathbf{L}} & \left\| \mathcal{W}\left (\mathbf{s} - \mathcal{FCB} \left(\mathbf{U}\mathbf{V}+\mathbf{L}\right)\right ) \right\|^2_2 \nonumber\\
 &+ \lambda \sum_{c=1}^{K}\text{TGV}^2\{U_c\}\:.
\end{eqnarray},
where $\mathbf{s}$ the measured data, $\mathcal{F}$ the non-uniform Fourier transform (NUFT) encoding operator, $\mathcal{C}$ the coil sensitivity operator, $\mathcal{B}$ the $B_0$ frequency shift operator and $\mathbf{L}$ represents the lipid signal ($N_{\mathbf{r}}$ by $T$ array) from skull that is reconstructed simultaneously with brain metabolite $\mathbf{U}$ and $\mathbf{V}$ but on a separate spatial support. $\mathcal{W}$, is a weighting operator of a Hamming window shape, and decreasing with the distance to the center of the $k$-space \cite{Ban2019,Klauser2021}. $\text{TGV}^2$ is the total generalized variation cost function with $\lambda$ the regularization parameter \cite{Knoll2011b}. The NUFT encoding operator of ECCENTRIC $\mathcal{F}$ is a discrete non-uniform Fourier transform of type 1:
\begin{equation}\label{eq:DFT}
(\mathcal{F}\rho)_{j,t} = \sum_{i}\mathrm{e}^{2\pi \mathrm{i} \mathbf{k}_j \cdot \mathbf{r}_i}\rho_{i,t}
\end{equation} 
with $\mathbf{r}_i$ are the uniform image space coordinates and ${\mathbf{k}_j}$ $k$-space sampling-point coordinates located on the 3D ECCENTRIC circles. 
The contamination of $\mathbf{U}$ and $\mathbf{V}$ by skull-lipid signal is prevented by filtering of the gradient descent during the reconstruction (Eq.\ref{eq:Recon}). Lipid signal is removed from each step of the gradient descent by applying the operator $(\mathbf{1}-\mathbb{P})$ with $\mathbb{P}$ the lipid subspace projection computed from the estimated lipid signal at the skull $\mathbf{L}$ \cite{Klauser2019}.

The regularization parameter used in the reconstruction was adjusted to $ \lambda =3 \times 10^{-4}$ by gradually increasing it from a low value until the noise-like artifacts in the metabolite maps disappeared \cite{Klauser2022,Knoll2011b}. The reconstruction rank, $K$, was determined qualitatively as the minimum number of components that contain some signal distinguishable from noise. For the 3D ECCENTRIC reconstruction, $K$ was specifically set to 40.

\nopagebreak[4]

\section{Methods}

\subsection{ECCENTRIC FID-MRSI acquisition parameters}
${}^1$H-FID-MRSI~\cite{Henning2009,Bogner2012} acquisition was implemented with 3D spherical stack-of-ECCENTRIC sampling as depicted in (Fig.\ref{fig:Sampling}{\it E}) on a 7T scanner (MAGNETOM Terra, Siemens Healthcare, Erlangen, Germany) running VE12U SP1 software and equipped with NOVA head coils (32Rx/1Tx and 32Rx/8Tx). The echo-time (TE) was set to the minimum possible: $0.9$~ms with a $27$~degree excitation flip-angle (FA) and $275$~ms repetition-time (TR). A slab selective excitation was performed with a Shinnar-LeRoux optimized pulse~\cite{Pauly1991,Klauser2021} with 6.5~kHz bandwidth and was preceded by four-pulses WET water suppression scheme~\cite{Ogg1994,Klauser2021} (Fig. 1). The FoV was $220\times 220\times 105$~mm${}^3$ (A-P/R-L/H-F) with $85$~mm-thick excited slab. A voxel size of $3.4\times 3.4\times 3.4$~mm${}^3$ (40.5 $\mu l$) was realized with a $64\times 64 \times 31$ matrix. The ECCENTRIC circles radius $R$ was set to $1/8\frac{n}{FoV}$ which corresponds to a diameter that encompasses a quarter of the width of the $k$-space, with $n$ being the square matrix size. With the chosen radius $R$, each ECCENTRIC circle sampled 51 points in $k$-space, enabling a spectral bandwidth of 2280~Hz without the need for temporal interleaving. The FID was sampled with 500~time-points, which resulted in a total FID duration of 220~ms. To obtain the fully sampled (AF=1) spherical 3D Stack-of-ECCENTRIC with these parameters, a total number of $N_c=4072$ circles is required, which corresponds to $18$~min~$40$~sec acquisition time (TA). For accelerated acquisitions, we decreased the number of circles to $N_c/AF$, with TA being shortened proportionally. For instance, with the same encoding parameters, AF=2 needs $N_c=2036$ in 9~min~20~sec, AF=3 needs $N_c=1357$ in 6~min~16~sec, AF=4 needs $N_c=1018$ in 4~min~40~sec, and so on. A very fast calibration scan of water reference data needed for coil combination, $B_1$ and $B_0$ field correction ($\mathcal{CB}$ operators in Eq.\ref{eq:Recon}), and image intensity normalization for comparable range of metabolite concentration values across subjects (in institutional units, I.U.) was acquired by turning off water suppression and using the same FoV, FA slab excitation, TR, FID duration and spectral bandwidth, with Rosette sampling at lower resolution ($23\times 23\times 19$) in $1$~min~$16$~sec.
The 3D-ECCENTRIC FID-MRSI data were processed by removing any residual water and then reconstructed using a TGV regularized model that also includes simultaneous suppression of scalp lipid signal (Eq. \ref{eq:Recon}).

The $B_0$  shimming of the 85~mm thick whole-brain slab was performed using the manufacturer methods that adjusted the shim currents over thirteen spherical harmonics coils: three 1st order, five 2nd order, and four 3rd order. The global linewidth of the water over the entire 85 mm slab was between 25-42~Hz across all subjects. In the majority of the subjects the global water linewidth was between 30-35~Hz. Adjustment of the $B_1+$ transmit and water suppression was subsequently performed with manufacturer methods. The entire adjustment procedure took between 1-2 min for every subject.

\subsection{Reconstruction of ECCENTRIC FID-MRSI metabolic images}

To obtain the metabolic images we employed a pipeline that included: 1) water removal by HSVD method  \cite{Barkhuijsen1987}, 2) CS-SENSE-LR reconstruction model from (Eq.\ref{eq:Recon}), and 3) spectral fitting by LCModel software \cite{Provencher1993}. Because FID gradient-echo excitation does not refocus chemical shift evolution during echo-time the spectra need first order phase correction, which was performed by backward linear prediction of the evolution \cite{Nassirpour2017}. A metabolite basis obtained by quantum mechanics simulations in GAMMA \cite{Smith1994} was utilized to fit and quantify twenty-one metabolites: phosphorylcholine (PCh), glycerophosphorylcholine (GPC), creatine (Cr), phosphocreatine (PCr), gamma-aminobutyric acid (GABA), glutamate (Glu), glutamine (Gln), glycine (Gly), glutathione (GSH), myo-inositol (Ins), N-acetylaspartate (NAA), N-acetyl aspartylglutamate (NAAG), scyllo-inositol (Sci), lactate (Lac), threonine (Thr), beta-glucose (bGlu), alanine (Ala), aspartate (Asp), ascorbate (Asc), serine (Ser), taurine (Tau). Phosphorylcholine and glycerophosphorylcholine were combined into total choline-containing compounds (Cho), while creatine and phosphocreatine were combined into total creatine (tCr). Concentration maps were then generated for the metabolites included in the simulated basis. 

The water reference signal was used as quantification reference by LCModel and the resulting concentration estimates were expressed in institutional units (I.U.). This allowed for comparisons of metabolite levels across both subjects and different metabolites. 
The ultra-short TE used in the ECCENTRIC MRSI data acquisition meant that $T_2$ relaxation correction was unnecessary for both metabolite and water signals. The results of the LCModel fitting for each voxel were further used to generate spatial maps of the concentration of each metabolite.
To assess the quality of the MRSI data and the goodness of fit, quality control maps of Cramer-Rao lower bounds (CRLB), line-width (FWHM), and SNR were generated from the LCModel fitting.

\subsection{ECCENTRIC FID-MRSI in high-resolution phantom}
Experimental performance of ECCENTRIC sampling was first tested on a high-resolution structural-metabolic phantom. We used a custom made phantom with geometry similar to Derenzo molecular imaging phantom \cite{Derenzo1977} containing 5 sets of tubes with diameters of $2$, $4$, $6$, $8$ and $10$~mm as shown in Fig.\ref{fig:Phantom}. Each set contained 6 tubes of identical diameter separated by a distance equal to twice the inner diameter positioned in a triangular configuration. In every set, the six tubes were filled with metabolite solutions containing 10~mM of creatine. Magnevist (Gd-DTPA) was added (1 mL/L) in each tube to shorten $T_1$ and create $T_1$-weighted contrast for structural MRI. The whole tube structure was inserted in a large cylindrical container (13.33~cm inner diameter) which was filled with 10~mM NaCl solution. Further details of phantom manufacturing and chemical composition are mentioned in \cite{Klauser2021}.

Due to the phantom's geometric structure being present only in the axial section, we opted for a 2D ECCENTRIC acquisition. The sampling scheme for 2D ECCENTRIC is identical to the central $k$-space partition used in 3D ECCENTRIC. The 2D ECCENTRIC acquisition maintained the same RF-pulse and FA but required slightly longer TE to $1.15$~ms and the TR was set to $450$~ms to accommodate the longer $T_1$ relaxation times in the phantom. The FID was measured with a spectral bandwidth of $2000$~Hz over $350$~ms, and successive acquisitions were performed with increased in-plane resolution. The circle radius ($R$) was set to $\frac{n}{8 FoV}$, $\frac{n}{8 FoV}$, $\frac{n}{9 FoV}$, and $\frac{n}{10 FoV}$ for $4.6$, $3.4$, $2.8$, and $2.0$~mm in-plane resolutions, respectively, to avoid temporal interleaving for any spatial resolution.

In addition to metabolite imaging, we conducted water imaging of the structural phantom at identical spatial resolutions to more comprehensively evaluate the spatial encoding performance of ECCENTRIC. To accomplish this, we employed the 2D ECCENTRIC sequence without water suppression, utilizing a short TR of $100$~ms and an FA of $40$~degrees. This choice was made to maximize the $T_1$-weighted ($T_1$w) contrast of the tubes within the cylindrical phantom. Subsequently, the first point of the acquired timeseries was reconstructed to generate the $T_1$-weighted water image. Both metabolite and water data were acquired with fully sampled ECCENTRIC. To investigate accelerations, the fully sampled ECCENTRIC data were retrospectively undersampled for acceleration factors (AF) between $2-12$. 

The effect of the acceleration on water and metabolite imaging was evaluated by analyzing the structural similarity index (SSIM) and correlation coefficient for all voxels inside the phantom with respect to the fully sampled data (Fig.\ref{fig:Phantom}).

\subsection{ECCENTRIC FID-MRSI in healthy volunteers}	
Five healthy volunteers were scanned at Athinoula A. Martinos Center For Biomedical Imaging for this study. The protocol was approved by the institutional ethics committee and written informed consent was given by all subjects before participation. The 3D ECCENTRIC FID-MRSI sequence described above was acquired with voxel size of $3.4$~mm isotropic at AF= 1,2,3 and 4 successively. In two volunteers the performance of 3D ECCENTRIC FID-MRSI was also tested at ultra-high resolution with voxel size of $2.5$~mm isotropic (matrix $88\times 88\times 43$, AF= 4, TA= 10~min~26~sec) and compared to $3.4$~mm isotropic (matrix $64\times 64\times 31$, AF= 2, TA= 9~min~20~sec). 

All volunteers were scanned with a $T_1$-weighted anatomical MP2RAGE sequence \cite{Marques2010} (1~mm isotropic, 4300 ms TR, 840~ms \& 2370~ms TI) for positioning of the MRSI FoV and for the generation of skull-masks that are needed for the lipid removal during the reconstruction and to exclude voxels located outside the head volume. 

\subsection{Quantitative analysis}	
For quantitative analysis of 3D ECCENTRIC FID-MRSI, the metabolite concentrations (I.U.) were  analyzed in each brain lobe and tissue type. This involved segmenting MP2RAGE images into gray matter, white matter, and cerebrospinal fluid using Freesurfer version 7.1.1 software \cite{Fischl2002}. Cerebral lobes were then identified utilizing a standard atlas, and a general linear model was employed to estimate metabolite concentrations within each atlas-defined structure \cite{Klauser2022}.

\subsection{Reproducibility analysis}	
The reproducibility of metabolite maps measured by 3D ECCENTRIC FID-MRSI were assessed by repeated imaging in four healthy volunteers. For the reproducibility analysis three data sets with $AF=3$ in each volunteer were compared.  One data set was acquired with $AF=3$, and the other two datasets were obtained by retrospective undersampling to $AF=3$ the data acquired with $AF=1$ and $AF=2$. Coefficients-of-variation (COV) were computed for individual anatomical regions from the three data sets. Both inter-measurement and inter-subject COVs were calculated and then averaged across subjects.

\section{Results}

\subsection{ECCENTRIC imaging in high-resolution phantom}

\begin{figure}[H]
\centering
\includegraphics[width=0.9\textwidth]{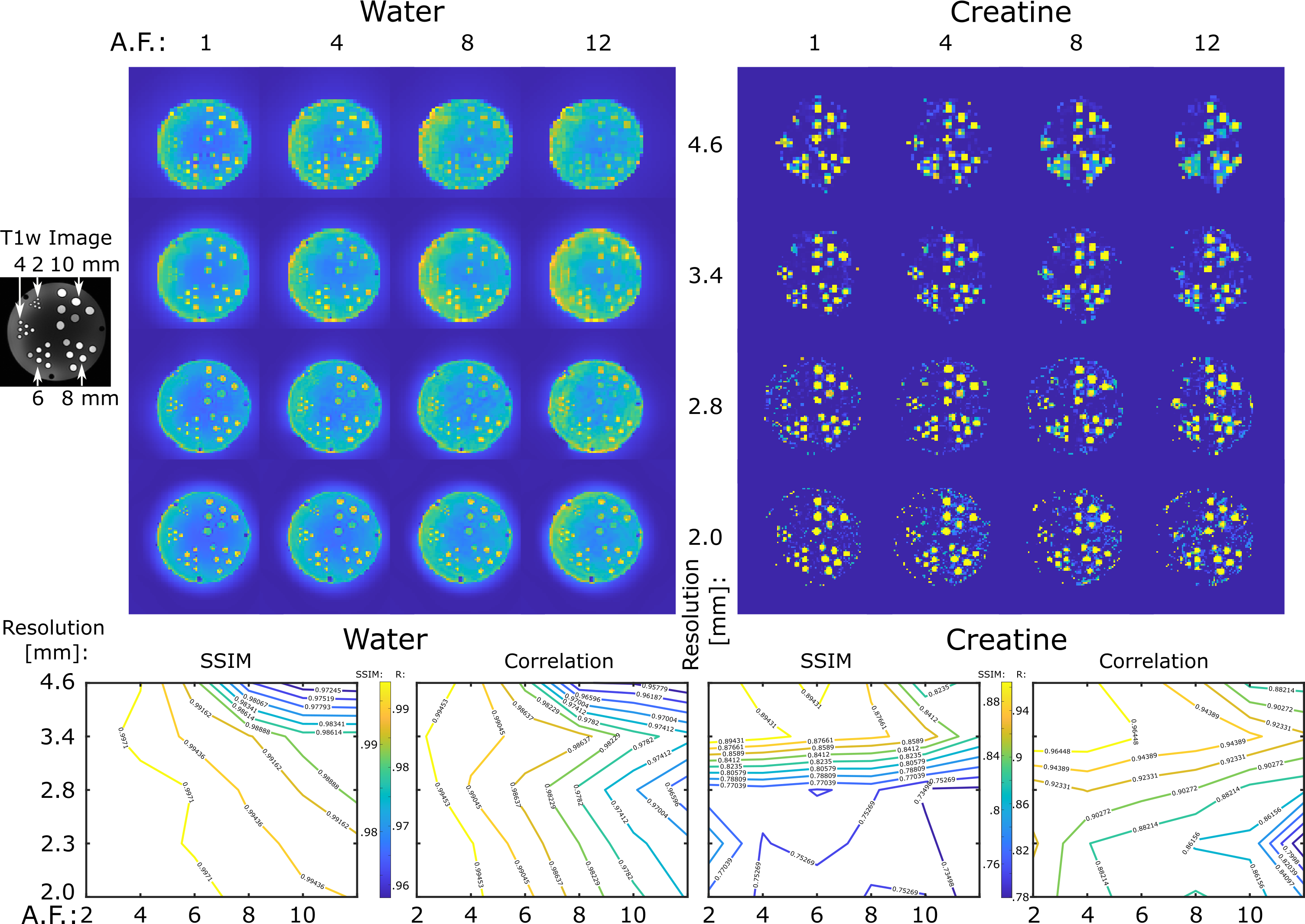}
\caption{ECCENTRIC imaging of water and metabolites in the high-resolution structural-metabolic phantom. ECCENTRIC performance was tested for spatial resolutions of 4.6, 3.4, 2.8, 2.0 ~mm and acceleration factors between 1-12. Top, examples of water and creatine images are shown for all 4 resolutions and 4 acceleration (AF: 1,4,8,12). Bottom, SSIM and correlation factors for each resolution and acceleration are calculated considering as ground truth the fully sampled image (AF=1).}
\label{fig:Phantom}
\end{figure}

In the series of water imaging experiments ECCENTRIC can resolve the structural details of the phantom up to the resolution targeted by the imaging protocol as can be seen in Figure\ref{fig:Phantom}. 

For water images no visible difference in image quality can be seen for retrospective accelerations factors up to AF=4, minor changes can be detected for AF between 4-8, and moderate loss of details for AF between 8-12 when compared to the fully sampled acquisition (AF=1). Considering AF=1 as ground truth, quantitative analysis reveals that SSIM $\ge$ 0.99 across all resolutions for accelerations up to AF = 4, and SSIM decreases to 0.97 for the highest acceleration and resolution tested. Similarly, correlation factors larger than 0.99 are observed up AF = 4, which decrease to 0.95 for the lowest resolution and largest acceleration factor. Comparing the different spatial resolutions, the higher CS accelerations show better performance for higher resolution images.

In the series of metabolite imaging experiments, ECCENTRIC was used to image the tCr metabolite present in the tubes with the the same resolutions and acceleration factors as in the water imaging (Fig.\ref{fig:Phantom}). These results of these experiments show that:
1) metabolite maps exhibit comparable quality for AF between 1-4, 2) for $AF>4$ there is  reduction in image details and an elevation of the noise level. Considering AF = 1 as ground truth, across the entire series of measurements SSIM range between 0.75-0.89, and correlation factors between 0.79-0.96. Unlike the water imaging results, the increasing acceleration does not yield higher SSIM and correlation factors as the spatial resolution increases. This discrepancy is likely attributed to the much lower ($10^3-10^4$ less) SNR of metabolites in comparison to water, which becomes critical for the smallest voxel size.

In particular, we note that for isotropic voxel size of 3.4 mm and for acceleration factors up to 4 we obtained the highest SSIM and correlation factors for metabolic imaging. Hence, we concluded that ECCENTRIC at 3.4 mm with AF between 1-4 represents a good starting protocol to further investigate fast high resolution metabolic imaging of the human brain.

\subsection{ECCENTRIC FID-MRSI in healthy volunteers}

Results from the high-resolution phantom imaging indicate that $AF$ up to 4 can be used for ECCENTRIC to obtain metabolic images that have very high agreement with the ground truth.
Guided by these results, we investigated the performance of ECCENTRIC for whole-brain metabolic imaging in healthy volunteers. 3D ECCENTRIC ${}^1$H-FID-MRSI was performed with a voxel size of 3.4~mm isotropic using $AF$ between 1 to 4. Our objective was to determine the feasibility of achieving a metabolic imaging protocol that delivers close to 3 mm isotropic whole-brain coverage in under 10 minutes using ECCENTRIC. This level of performance is comparable to other advanced MR imaging methods, such as CEST and perfusion imaging.

\begin{figure}[t]
\centering
\includegraphics[width=0.9\textwidth]{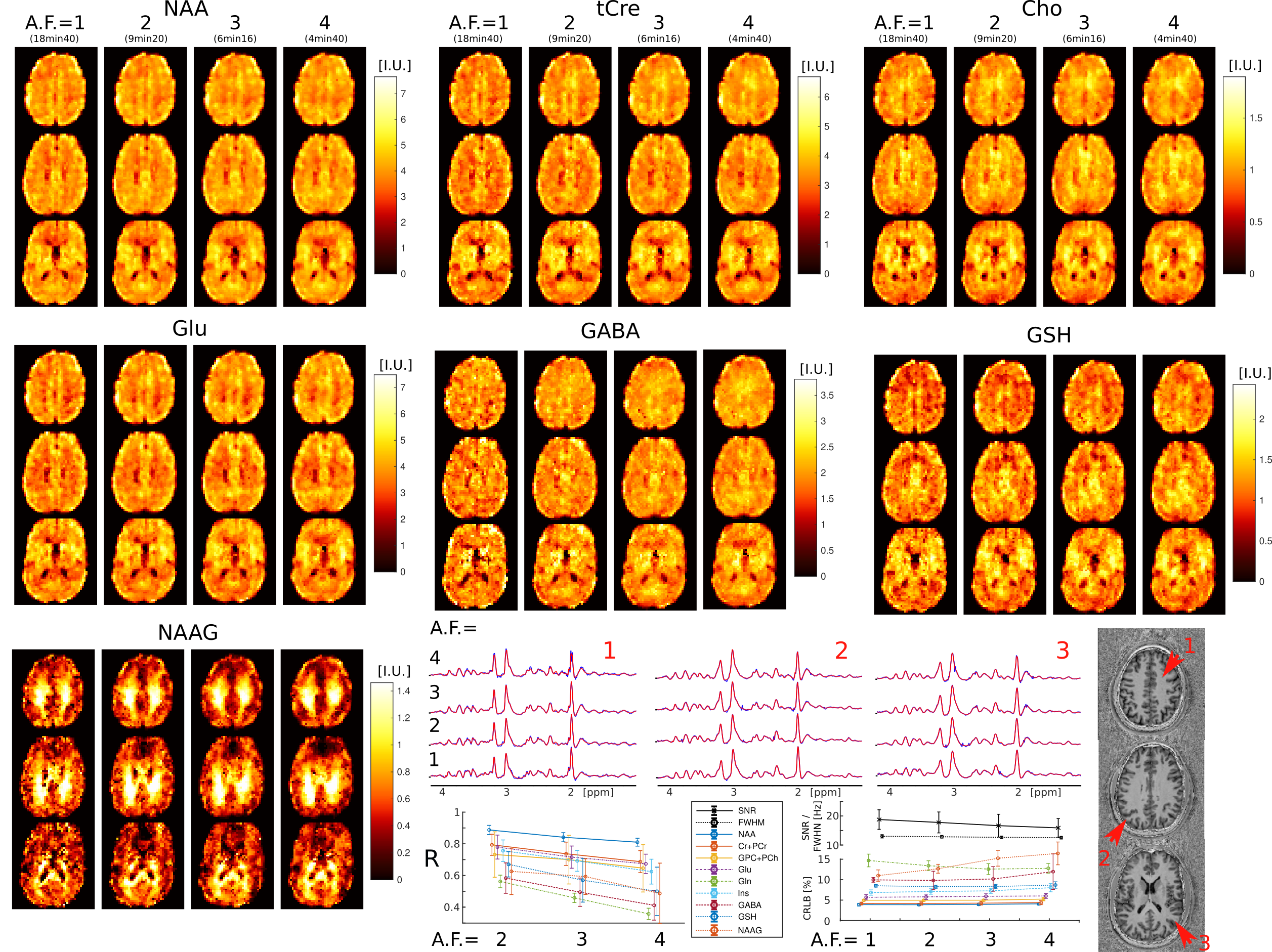}
\caption{3D ECCENTRIC ${}^1$H-FID-MRSI metabolic images of human brain acquired in a healthy volunteer with $3.4$~mm isotropic voxel size and CS acceleration factors AF=1-4. Top, metabolite maps of seven relevant brain metabolites (NAA, tCre, Cho, Glu, GABA, GSH, and NAAG) are shown for all acceleration factors (AF). Spectra from three brain locations indicated by red arrows on the anatomical image. At the bottom, the left plot displays the correlation coefficients between accelerated images ($AF=2,3,4$) and fully sampled images ($AF=1$), while the right plot show the LCModel quantification error (CRLB), linewidth (FWHM), and SNR.}
\label{fig:MetabVolAcc}
\end{figure}

Examples of metabolic images for seven metabolites obtained with retrospective $AF = 1-4$ are shown in Fig.\ref{fig:MetabVolAcc}. Very similar structural details and tissue contrast of metabolic images are obtained for all accelerations compared to the fully sampled data. This is visible also by inspecting spectra that show the same metabolic profile across acceleration factors ($AF=1-4$). The CS accelerations ($AF=2, 3, 4$) were achieved by retrospectively undersampling the fully acquired data ($AF=1$). The purpose of this was to focus the analysis on the effects of CS acceleration, while avoiding any image differences that could be caused by head motion during different acquisitions.

Visual inspection of metabolic images reveal that: 1) tCre, Glu and GABA have larger signal in gray matter than white matter, 2) NAA has more signal in gray than white matter, but with lower gray-white matter contrast compared to tCre, Glu and GABA, 3) Cho has higher signal in frontal white matter than gray matter, 4) NAAG has the largest contrast from all metabolites, with much larger signal in white matter compared to gray matter. Metabolic images obtained with $AF = 1$  and $AF = 2$ are largely identical. Minor blurring of fine structural details starts to become noticeable for $AF\geq 3$, however adequate delineation of gray-white matter folding is maintained up to $AF = 4$. Additionally, the performance of ECCENTRIC to reveal brain structural details was probed also by water imaging in four healthy volunteers. Brain water imaging by ECCENTRIC shows comparable performance for $AF$ between 1-10 (Supplementary Material Fig.S1), similar to the phantom results .

Quantitative image analysis shows that the correlation between the accelerated and fully sampled metabolic images is high (R $>$ 0.7) for metabolites that have a high SNR ($>10$) such as NAA, Cho, tCr, Glu and Ins, while metabolites of lower SNR such as Gln, GABA, GSH, and NAAG exhibit lower correlations (R = $0.4-0.7$). The error (CRLB) of spectral fitting is below 20$\%$, which indicates very high goodness-of-fit by LCModel software. The CRLB does not degrades with the acceleration factor, except for NAAG and GABA, although even in this case it does not exceeds the 20$\%$ limit. The SNR shows only  a minor decrease between $AF=1$ (SNR=17) and $AF=4$ (SNR=15), while the linewidth does not degrade with acceleration.

\subsection{Quantitative analysis of ECCENTRIC metabolic imaging}	

Table \ref{table:CRLB_SNR_FWHM_Conc_Healthy} presents brain regional concentrations of nine metabolites quantified using water signal as reference and expressed in institutional units (I.U.). The concentrations were calculated in the gray and white matter of the five major brain lobes across the five healthy volunteers. Our results indicate that: 1) six metabolites have higher concentrations in gray-matter compared to white matter (GM/WM = 1.19 tCre, 1.12 NAA, 1.2 Glu, 1.34 Gln, 1.15 GABA, 1.11 GSH); 2) two metabolites have higher concentrations in white-matter compared to gray-matter (GM/WM = 0.96 Cho, 0.47 NAAG); 3) one metabolite has region dependent gray/white-matter ratio (GM/WM = 1.17-0.87 Ins). The largest gray-white matter contrast is exhibited by NAAG due to its specific compartmentalization in white matter. 

Quantitative parameters for the quality of MRSI data are also listed in Table \ref{table:CRLB_SNR_FWHM_Conc_Healthy}, including the precision of metabolite quantification by the Cramer-Rao lower bounds (CRLB), the SNR and spectral linewidth (FWHM). It can be seen that mean CRLB is below $<20\%$ for all the metabolites across the imaged whole-brain volume. In particular, mean CRLB is below $<6\%$ for the five metabolites with highest SNR (NAA, tCre, Cho, Glu and Ins), between $8\%$-$10\%$ for two metabolites (GABA and GSH) and between $14\%$-$20\%$ for other two metabolites (Glu and NAAG). Across the brain the mean SNR is larger than 20 and the mean linewidth is less than 12 Hz (0.04 ppm).


\begin{table}[H]
\centering
 \resizebox{\textwidth}{!}{%
 \renewcommand{\arraystretch}{1.5}
 \begin{tabular}{||c c c c c c c c c c c|c||} 
 \hline
 Mean across volunteers & \multicolumn{2}{c}{Frontal} & \multicolumn{2}{c}{Limbic} & \multicolumn{2}{c}{Parietal} & \multicolumn{2}{c}{Occipital} & \multicolumn{2}{c|}{Temporal} & usable voxels\\
 (Standard deviation) & WM & GM & WM & GM & WM & GM & WM & GM & WM & GM & mean \% (std)\\ [0.5ex] 
 \hline
 tCre [I.U.] & 3.34 ( 0.34 )& 3.84 ( 0.04 )& 3.22 ( 0.35 )& 4.46 ( 0.28 )& 3.30 ( 0.28 )& 3.95 ( 0.28 )& 3.31 ( 0.30 )& 3.39 ( 0.19 )& 3.09 ( 0.25 )& 3.81 ( 0.19 )& \\ 
  tCre CRLB [\%] & 3.25 ( 0.24 )	& 3.54 ( 0.41 )	& 3.38 ( 0.31 )	& 3.18 ( 0.30 )	& 3.21 ( 0.33 )	& 3.26 ( 0.26 )	& 3.45 ( 0.62 )	& 3.64 ( 0.62 )	& 3.42 ( 0.37 )	& 3.33 ( 0.26 ) & 73.3 (3.3)\\ 
 \hline
  NAA  [I.U.]	& 4.16 ( 0.98 )& 4.51 ( 0.66 )& 4.07 ( 1.01 )& 5.38 ( 0.90 )& 4.31 ( 0.96 )& 4.84 ( 0.88 )& 4.33 ( 0.98 )& 4.18 ( 0.85 )& 3.91 ( 0.84 )& 4.35 ( 0.73 )& \\ 
 NAA CRLB [\%] & 3.13 ( 0.34 )	& 3.55 ( 0.64 )	& 3.26 ( 0.33 )	& 3.07 ( 0.36 )	& 2.86 ( 0.26 )	& 3.05 ( 0.37 )	& 3.30 ( 0.54 )	& 3.63 ( 0.64 )	& 3.32 ( 0.39 )	& 3.32 ( 0.34 ) & 73.0 (3.5)\\ 
 \hline 
 Ins [I.U.]& 3.49 ( 0.52 )& 3.51 ( 0.35 )& 3.65 ( 0.60 )& 4.28 ( 0.48 )& 3.73 ( 0.66 )& 3.58 ( 0.29 )& 3.74 ( 0.65 )& 3.27 ( 0.57 )& 3.56 ( 0.65 )& 3.39 ( 0.51 ) & \\
 Ins CRLB [\%] & 5.33 ( 0.64 ) & 6.04 ( 0.94 ) & 5.29 ( 0.52 ) &	 
5.36 ( 0.42 )	& 5.16 ( 0.42 ) & 5.63 ( 0.52 ) & 5.38 ( 0.48 ) & 5.84 ( 0.50 )	& 5.34 ( 0.33 ) & 5.68 ( 0.40 ) & 73.0 (3.5) \\
 \hline
 GPC+PCh [I.U.]& 1.03 ( 0.10 )& 0.93 ( 0.05 )& 1.10 ( 0.10 )& 1.24 ( 0.08 )& 1.06 ( 0.11 )& 0.95 ( 0.09 )& 0.94 ( 0.12 )& 0.79 ( 0.06 )& 0.99 ( 0.07 )& 0.98 ( 0.12 ) & \\ 
 GPC+PCh CRLB [\%] & 3.63 ( 0.34 )	& 4.24 ( 0.59 )	& 3.46 ( 0.31 )	& 3.56 ( 0.27 )	& 3.56 ( 0.29 )	& 3.97 ( 0.33 )	& 4.13 ( 0.57 )	& 4.59 ( 0.64 )	& 3.68 ( 0.29 )	& 3.94 ( 0.25 ) & 73.5 (3.3) \\
 \hline
 Glu [I.U.]& 3.86 ( 0.53 )& 4.50 ( 0.29 )& 3.72 ( 0.49 )& 5.24 ( 0.46 )& 3.86 ( 0.39 )& 4.61 ( 0.39 )& 3.80 ( 0.36 )& 3.82 ( 0.38 )& 3.50 ( 0.35 )& 4.29 ( 0.44 ) & \\
 Glu CRLB [\%] &
 4.51 ( 0.51 )	& 5.01 ( 0.85 )	& 4.72 ( 0.59 )	& 4.37 ( 0.47 )	& 4.31 ( 0.53 )	& 4.48 ( 0.55 )	& 4.82 ( 1.14 )	& 5.22 ( 1.32 )	& 4.80 ( 0.74 )	& 4.71 ( 0.64 ) & 72.8 (3.8) \\
 \hline
 Gln  [I.U.]& 0.89 ( 0.06 )& 1.24 ( 0.25 )& 0.83 ( 0.06 )& 1.33 ( 0.16 )& 0.80 ( 0.16 )& 1.08 ( 0.26 )& 0.87 ( 0.22 )& 0.91 ( 0.21 )& 0.80 ( 0.09 )& 1.07 ( 0.22 ) & \\
 Gln CRLB [\%] & 15.93 ( 2.45 )	& 14.81 ( 2.22 )	& 17.72 ( 2.68 )	& 14.38 ( 2.48 )	& 17.92 ( 4.78 )	& 16.22 ( 3.24 )	& 18.99 ( 5.14 )	& 18.77 ( 4.11 )	& 18.30 ( 2.95 )	& 16.45 ( 2.60 ) & 56.3 (4.6) \\ 
 \hline
 GABA [I.U.]& 1.34 ( 0.09 )& 1.48 ( 0.29 )& 1.34 ( 0.10 )& 1.79 ( 0.37 )& 1.37 ( 0.14 )& 1.58 ( 0.30 )& 1.33 ( 0.18 )& 1.37 ( 0.18 )& 1.21 ( 0.16 )& 1.39 ( 0.32 ) & \\
 GABA CRLB [\%] & 8.68 ( 1.10 )	& 9.61 ( 1.51 )	& 8.86 ( 1.50 )	& 8.52 ( 1.74 )	& 8.54 ( 2.08 )	& 8.78 ( 1.76 )	& 9.48 ( 2.34 )	& 9.90 ( 2.22 )	& 9.60 ( 2.16 )	& 9.52 ( 2.09 ) & 68.3 (1.5) \\ 
 \hline
  GSH [I.U.]& 1.07 ( 0.24 )& 1.22 ( 0.05 )& 1.11 ( 0.26 )& 1.45 ( 0.20 )& 1.10 ( 0.20 )& 1.17 ( 0.07 )& 1.05 ( 0.18 )& 1.00 ( 0.09 )& 1.02 ( 0.17 )& 1.14 ( 0.12 ) & \\
 GSH CRLB [\%] & 8.20 ( 0.88 )	& 8.77 ( 1.15 )	& 8.29 ( 0.93 )	& 7.96 ( 0.79 )	& 8.04 ( 1.00 )	& 8.43 ( 0.87 )	& 9.12 ( 2.01 )	& 9.80 ( 1.90 )	& 8.65 ( 1.23 )	& 8.81 ( 1.17 ) & 71.5 (3.1) \\
 \hline
 NAAG [I.U.] & 0.82 ( 0.14 )& 0.39 ( 0.12 )& 0.90 ( 0.13 )& 0.49 ( 0.17 )& 0.96 ( 0.17 )& 0.34 ( 0.13 )& 0.78 ( 0.16 )& 0.41 ( 0.12 )& 0.76 ( 0.11 )& 0.35 ( 0.10 ) & \\
 NAAG CRLB [\%] & 14.11 ( 4.10 )	& 19.02 ( 5.07 )	& 12.43 ( 3.87 )	& 17.35 ( 6.07 )	& 12.01 ( 3.14 )	& 18.03 ( 5.15 )	& 15.51 ( 5.08 )	& 19.40 ( 6.62 )	& 15.36 ( 5.01 )	& 20.13 ( 6.40 ) & 45.3 (11.6) \\ 
 \hline
 SNR & 24.64 ( 4.41 )	& 23.67 ( 3.84 )	& 22.92 ( 4.50 )	& 24.53 ( 4.02 )	& 25.64 ( 4.82 )	& 24.73 ( 4.17 )	& 22.48 ( 6.13 )	& 21.32 ( 5.56 )	& 22.07 ( 4.23 )	& 22.40 ( 3.37 ) & $\emptyset$\\ 
 \hline
 FWHM [Hz] & 11.43 ( 0.91 )	& 12.09 ( 0.77 )	& 11.28 ( 1.15 )	& 11.13 ( 1.13 )	& 10.21 ( 0.79 )	& 11.02 ( 0.84 )	& 11.64 ( 0.91 )	& 12.33 ( 1.04 )	& 12.26 ( 1.23 )	& 12.38 ( 1.22 ) & $\emptyset $\\

 \hline\hline
 \end{tabular} 
 }
 \caption{Metabolite concentrations (I.U.) and quantification error (Cramér-Rao lower bound, CRLB \%) in each brain lobe and tissue type.  The two bottom rows present the SNR and FWHM values. The last column shows the percentage of voxels inside the brain and FoV that meet the criteria of good quality: CRLB $< 20\%$, FWHM $< 0.07ppm$, SNR $>5$. The values are calculated as the average (standard deviation) across the healthy volunteers imaged by 3D ECCENTRCIC at 3.4~mm isotropic with $AF=2$ (9min:20s acquisition time). }
 \label{table:CRLB_SNR_FWHM_Conc_Healthy}
\end{table}


\subsection{Reproducibility of ECCENTRIC metabolic imaging}	

\begin{figure}[H]
\centering
\includegraphics[width=0.75\textwidth]{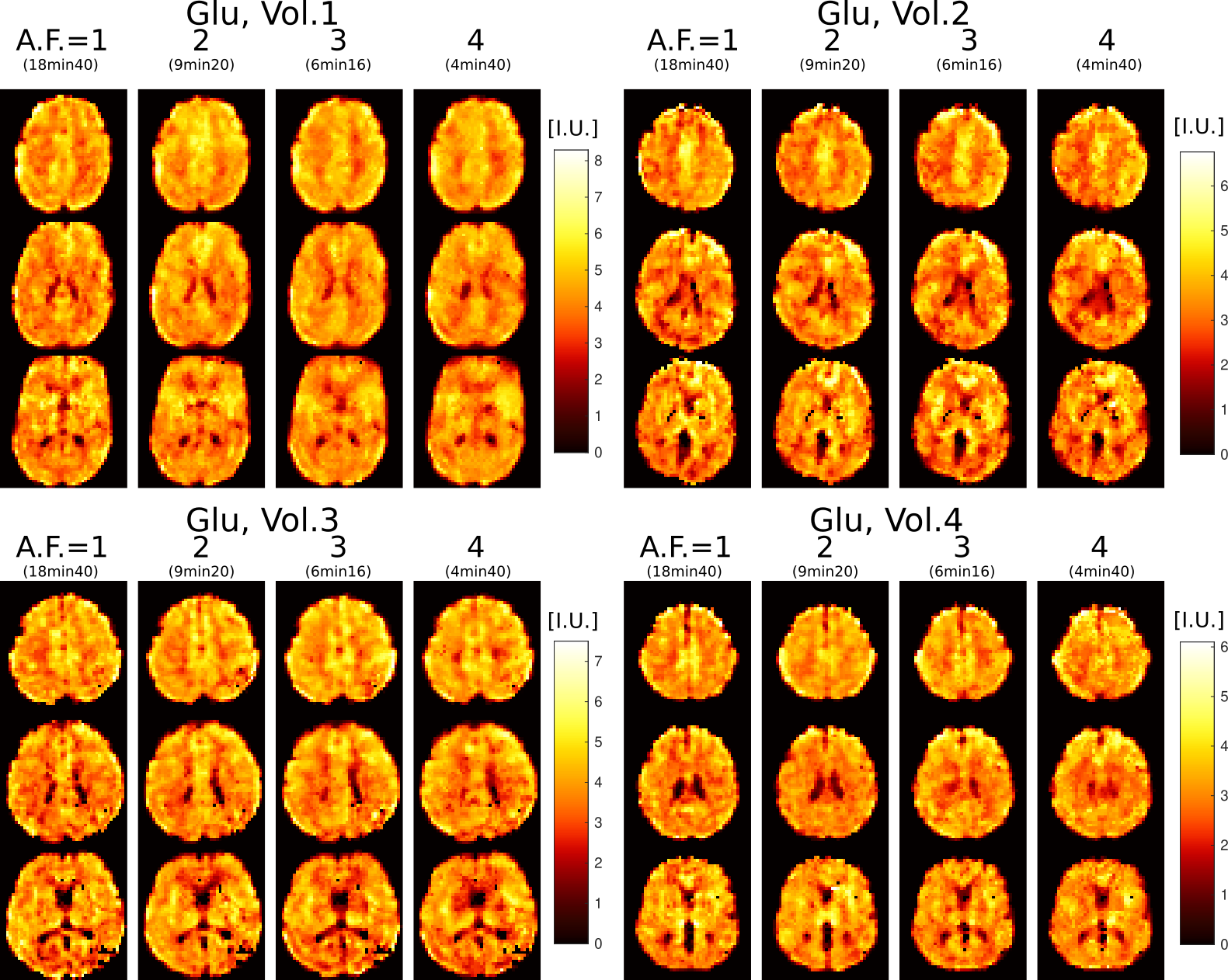}
\caption{Glutamate imaging at $3.4$mm isotropic voxel size in four healthy volunteers scanned with 3D ECCENTRIC FID-MRSI in four successive acquisitions with increasing accelerations AF=1, 2, 3 and 4. Three slices are shown for each volunteer at each acceleration.}
\label{fig:MetabGluAcc}
\end{figure}

Results from repeated triplicate measurements are shown in Fig.\ref{fig:MetabGluAcc} for Glu imaging. Due to high concentration of Glu in gray mater, Glu images have high gray-white matter contrast and show fine structural details of brain that can be used to visually assess the stability and reproducibility of test-retest imaging. It can be seen that across all four scans in all four subjects the metabolite images appear visually similar. We note that with repeated measurements some anatomical differences may also be attributed to slight head motion.  

The inter-measurement COV for mapping metabolite concentrations across brain regions are presented in Table \ref{SDtable:COV} and the inter-subject COV in Table  \ref{SDtable:IV-COV}. Inter-measurement COV smaller than $7\%$ are observed for five metabolites (NAA, creatine, myo-inositol, choline, glutamate) that are the most abundant in the brain. COV between $8\%-14\%$ are obtained for glutamine, glutathione and GABA. NAAG has higher COV in brain regions (gray matter) where its concentration is low. The inter-subject COV in (Table \ref{SDtable:IV-COV}) reflects the biological variability of metabolite concentrations in the gray and white matter across the group of subjects.
 

\begin{table}[!tb]
\centering
\begin{center}
 \renewcommand{\arraystretch}{1.5}
 \begin{tabular}{||c c c c c c c c c c c||} 
  \hline
  & \multicolumn{10}{c|}{Inter-measurement COV } \\
  & \multicolumn{2}{c}{Frontal} & \multicolumn{2}{c}{Limbic} & \multicolumn{2}{c}{Parietal} & \multicolumn{2}{c}{Occipital} & \multicolumn{2}{c||}{Temporal} \\
  & WM & GM & WM & GM & WM & GM & WM & GM & WM & GM\\ [0.5ex] 
  \hline
  NAA & 0.06 & 0.03 & 0.06 & 0.05 & 0.06 & 0.05 & 0.05 & 0.07 & 0.06 & 0.06 \\ \hline
tCr & 0.05 & 0.05 & 0.05 & 0.03 & 0.03 & 0.04 & 0.04 & 0.03 & 0.06 & 0.05 \\ \hline
Ins & 0.05 & 0.04 & 0.07 & 0.06 & 0.05 & 0.05 & 0.04 & 0.06 & 0.06 & 0.07 \\ \hline
GPC+PCh & 0.04 & 0.03 & 0.03 & 0.04 & 0.04 & 0.06 & 0.03 & 0.04 & 0.03 & 0.05 \\ \hline
Glu & 0.05 & 0.03 & 0.05 & 0.05 & 0.07 & 0.07 & 0.07 & 0.07 & 0.06 & 0.07 \\ \hline
Gln & 0.14 & 0.10 & 0.15 & 0.11 & 0.12 & 0.12 & 0.12 & 0.07 & 0.14 & 0.08 \\ \hline
GABA & 0.07 & 0.11 & 0.05 & 0.07 & 0.09 & 0.12 & 0.05 & 0.11 & 0.06 & 0.11 \\ \hline
GSH & 0.12 & 0.10 & 0.14 & 0.13 & 0.13 & 0.13 & 0.13 & 0.10 & 0.14 & 0.11 \\ \hline
NAAG & 0.23 & 0.41 & 0.28 & 0.33 & 0.24 & 0.55 & 0.26 & 0.40 & 0.28 & 0.54 \\ \hline
  \hline
 \end{tabular} 
\end{center}
 \caption{ The inter-measurement coefficient of variation (COV) for each metabolite determined in every lobe and tissue type for the 3D ECCENTRIC FID-MRSI acquired at 3.4 mm isotropic resolution in 6min:16sec (AF=3). 
 }
 \label{SDtable:COV}
\end{table}

\begin{table}[H]
\centering
    \begin{center}
   \renewcommand{\arraystretch}{1.5}
    \begin{tabular}{||c c c c c c c c c c c||} 
        \hline
        & \multicolumn{10}{c|}{Inter-subject COV } \\
       & \multicolumn{2}{c}{Frontal} & \multicolumn{2}{c}{Limbic} & \multicolumn{2}{c}{Parietal} & \multicolumn{2}{c}{Occipital} & \multicolumn{2}{c||}{Temporal} \\
         &   WM &  GM  &   WM &   GM &  WM &  GM &   WM &  GM &  WM & GM\\ [0.5ex] 
        \hline
      NAA & 0.21  & 0.15 & 0.23 & 0.17 & 0.23 & 0.17 & 0.24 & 0.23 & 0.23 & 0.22 \\
Cr+PCr & 0.10 & 0.05 & 0.12 & 0.07 & 0.11 & 0.05 & 0.09 & 0.09 & 0.11 & 0.09 \\
Ins & 0.18 & 0.15 & 0.22 & 0.17 & 0.22 & 0.14 & 0.21 & 0.26 &,0.24 & 0.19 \\
GPC+PCh & 0.06 & 0.07 & 0.07 & 0.06 & 0.10 & 0.10 & 0.12 & 0.14 & 0.11 & 0.15 \\
Glu & 0.19 & 0.13 & 0.20 & 0.16 & 0.20 & 0.15 & 0.18 & 0.18 & 0.18 & 0.20 \\
Gln & 0.16 & 0.25 & 0.12 & 0.16 & 0.18 & 0.21 & 0.21 & 0.23 & 0.17 & 0.18 \\
GABA & 0.37 & 0.33 & 0.39 & 0.38 & 0.36 & 0.36 & 0.38 & 0.39 & 0.37 & 0.40 \\
GSH & 0.13 & 0.09 & 0.15 & 0.11 & 0.13 & 0.07 & 0.11  & 0.13 & 0.11 & 0.12 \\
NAAG & 0.32 & 0.37 & 0.35 & 0.29 & 0.45 & 0.47 & 0.50 & 0.48 & 0.38 & 0.29 \\ \hline
        \hline
    \end{tabular} 
\end{center}
    \caption{The inter-subject coefficient of variation (COV) for each metabolite determined in every lobe and tissue type for the 3D ECCENTRIC FID-MRSI acquired at 3.4 mm isotropic resolution in 6min:16sec (AF=3). 
    }
    \label{SDtable:IV-COV}
\end{table}

The results from Tables \ref{SDtable:COV} and \ref{SDtable:IV-COV} show that 3D ECCENTRIC FID-MRSI had reproducible and stable performance in three repeat measurements. The variability due to the technical performance (inter-measurement COV in Table \ref{SDtable:COV}) of 3D ECCENTRIC FID-MRSI is much lower (2-4 times smaller) than the biological variability (inter-subject COV in Table \ref{SDtable:IV-COV}) of brain metabolism across young healthy individuals of close age. The low variability of ECCENTRIC in repeat measurements indicates high precision of metabolite quantification and high potential for longitudinal studies to detect metabolite changes due to disease, treatment and functional tests. The quantification of the five main metabolites that have the highest SNR in brain MRSI (NAA, tCre, Cho, Ins, Glu) shows the lowest variability, with a slight increase in the case of less abundant metabolites (Gln, GABA and GSH). The highest variability is noticed for NAAG outside of the fronto-parietal white matter due to its specific localization in this brain area. We note that in-vivo variability of metabolite quantification in repeat measurements is also influenced by patient motion and scanner stability in addition to 3D-ECCENTRIC MRSI, hence methods that reduce the effects of motion and field drift \cite{Bogner2014,Andronesi2021} are likely to reduce variability.

\nopagebreak[4]
\subsection{Ultra-high resolution metabolic imaging of human brain using 3D ECCENTRIC FID-MRSI}

\begin{figure}[H]
\centering
\includegraphics[width=0.75\textwidth]{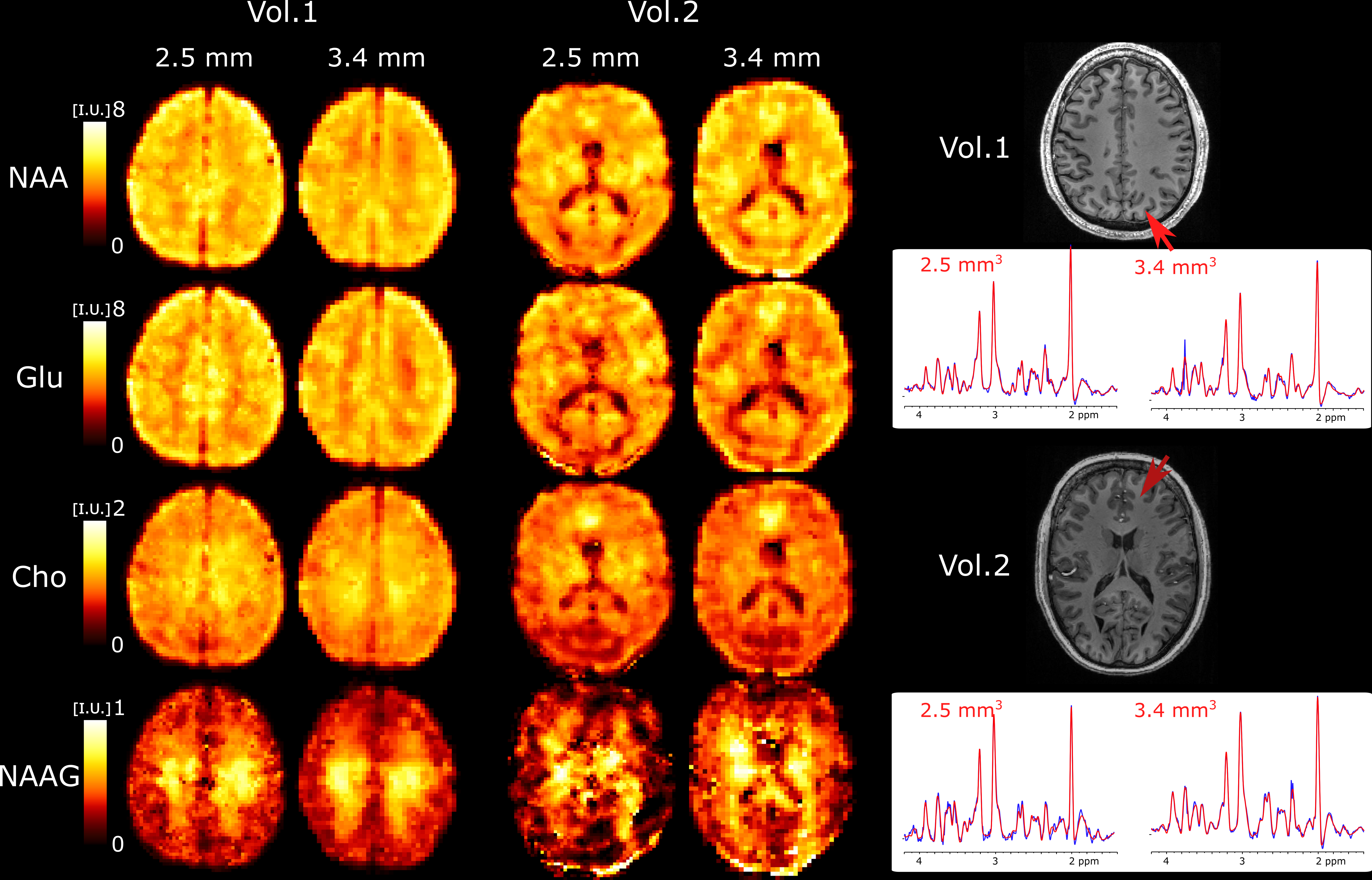}
\caption{Ultra-high resolution metabolic imaging acquired with 3D ECCENTRIC FID-MRSI at $2.5$~mm isotropic voxel size ($AF = 4$, TA=$10min:26s$) in two healthy volunteers. The ultra-high resolution metabolic imaging is compared to 3D ECCENTRIC FID-MRSI at the typical voxel size of $3.4$~mm isotropic ($AF=2$, TA=$9min:20s$). Right, two spectra from both spatial resolution and corresponding to the red arrow location are shown. The blue line represents the MRSI data and the red line is the fit performed by LCModel.}
\label{fig:HRPats}
\end{figure}

We further explored the performance of 3D ECCENTRIC FID-MRSI for ultra-high resolution metabolic imaging in several healthy volunteers. Based on the high SNR of the 3.4 mm data we expected that smaller voxels at higher resolution will still provide sufficient SNR for metabolite imaging. Fig.\ref{fig:HRPats} shows metabolic images obtained using 3D ECCENTRIC FID-MRSI with isotropic voxel size of 2.5~mm in two healthy volunteers. 

To achieve a feasible scan time, we used CS with $AF=4$. We demonstrated at the beginning of our work that $AF=4$ provides metabolic maps that are similar to those obtained through fully sampled 3D ECCENTRIC FID-MRSI. The $AF=4$ acceleration enabled the acquisition of 3D ECCENTRIC FID-MRSI at $2.5$~mm isotropic resolution in 10min:26s. For comparison we also acquired the typical 3D ECCENTRIC FID-MRSI at $3.4$~mm isotropic with $AF=2$ acceleration in 9min:20s. As readily apparent by visual inspection, the metabolic maps at higher spatial resolution provide sharper delineation of the brain structure. No compromise is visible for signal-to-noise, contrast-to noise or other data quality metric at ultra-high resolution compared to typical resolution. We note that the acquisition time of 3D ECCENTRIC FID-MRSI at $2.5$~mm with $AF=4$ is only slightly longer ($1$~min) than at $3.4$~mm with $AF=2$. However, for the same acceleration factor the acquisition time of 3D ECCENTRIC FID-MRSI at $3.4$~mm is $2.2$ times faster than at $2.5$~mm. 

\nopagebreak[4]
\section{Discussion}
Our results here demonstrate that 3D ECCENTRIC ${}^1$H-FID-MRSI at 7T can simultaneously image an extended neurochemical panel of 10-14 metabolites with high SNR at high spatial resolution across whole-brain and with acquisition times that are feasible for human imaging. Particularly, we showed that the acquisition of fast non-Cartesian MRSI can be further accelerated up to 4-fold by CS, allowing metabolic imaging at 3.4~mm isotropic resolution in 4min:40s and at 2.5~mm isotropic resolution in 10min:26s, respectively. The CS-SENSE-LR reconstruction produces metabolic images with an effective voxel size identical to the nominal size \cite{Klauser2021}. This provides an advantage compared to other filtered reconstructions \cite{Hingerl2019,Hangel2020} which increase the effective voxel volume. ECCENTRIC preserves the features of metabolic images across accelerations. When accelerated up to 4-fold by compressed sensing the loss of image quality is minor and the metabolic images effectively visualize the laminar structure of the brain similar to the unaccelerated (AF=1) ground truth. By design ECCENTRIC acquisition and reconstruction preserves the SNR at high accelerations. This is achieved through the full sampling of the center of k-space and the use of a low-rank denoising reconstruction.

Here we investigated the performance of 3D ECCENTRIC ${}^1$H-FID-MRSI for two applications scenarios: 1) high resolution metabolic imaging (3.4~mm in 4min:40s) for studies that need to minimize imaging time, and 2) ultra-high resolution metabolic imaging (2.5~mm in 10min:26s) for applications that need to probe brain neurochemistry with highest structural detail. Both of these protocols, represent a significant advancement for non-invasive imaging of human brain metabolism by \textit{in vivo} MRSI. 

Results obtained with the 3.4~mm imaging protocol show good delineation of brain structures. At 2.5~mm ultra-high resolution there is increased gray-white matter contrast of metabolites due to less partial volume effect which reveals the brain folding more clearly than at 3.4~mm. Several metabolites show particularly high contrast between gray and white matter in healthy brain, such as the energy buffer tCre, the neurotransmitter Glu and the dipeptide NAAG. In particular, NAAG is the most abundant dipeptide in the brain, which is selectively localized in several regions \cite{Pouwels1998} where it neuromodulates the glutamatergic synapses required for normal brain activity. Importantly, NAAG is also implicated in neurodegenerative diseases, schizophrenia, stroke, epilepsy, traumatic brain injury and pain \cite{Morland2022}. Our data show the highest resolution of 3D imaging for NAAG to date. 3D ECCENTRIC FID-MRSI provides high quality images of the NAAG brain distribution, which can be used to answer important questions in basic and clinical neurosciences. Good quality metabolic images are obtained also for some of the most important but challenging metabolites such as GABA, Gln, and GSH. The combination of higher SNR and narrower linewidth (FWHM) results in lower CRLB for GABA, Gln, and GSH. The potential of short-echo FID spectra to detect GABA, Gln, and GSH at ultra-high field is supported also by previous findings reported in 9.4T studies \cite{Nassirpour2017,Ziegs2023}. Here, we extend the imaging of NAAG, GABA, Gln and GSH from single-slice to whole-brain and show that this is feasible at 7T which is more available for ultra-high field human imaging compared to 9.4T. Due to FDA approval the 7T imaging has reached clinical use where ECCENTRIC is expected to have a great contribution.

While correlation coefficients between the ground truth and accelerated ECCENTRIC acquisitions generally exhibit lower values for metabolic imaging (Fig.\ref{fig:MetabVolAcc}) compared to water imaging (Fig. \ref{fig:Phantom} \& Suppl. Figs. S1), visual assessments indicate that the quality of metabolite mapping in the healthy brain is consistently preserved across all acceleration levels. Notably, the discrepancy between the results for metabolites and water images arises from the fact that, unlike water images where correlation coefficients are directly determined from reconstructed images, the correlation coefficients for metabolic images are influenced by additional processing steps such as water removal, fat removal, and LCModel fitting. These additional processing steps of MRSI data introduce variability that contributes to the lower correlation coefficients observed in metabolic imaging results.

The 3D ECCENTRIC FID-MRSI showed robust performance in healthy volunteers. The high quality of the data was achieved through the use of third-order shimming, which provides more uniform $B_0$ field across the brain, as well as the shortened scan time, which minimized the scanner drift and possibility of subject motion. The scanner drift typically ranged from 5-10~Hz over a 10min scan time.

ECCENTRIC encoding is highly versatile with flexible choice of FoV, spatial resolution, spectral bandwidth that can be set to optimize SNR and acquisition time. The advantage and strength of ECCENTRIC is enabled by the possibility to freely choose the radius and position of circle trajectories in covering the k-space: 1) the free choice of circle radius allows freedom in setting FoV, spatial resolution and spectral bandwidth without the need of temporal interleaving, 2) the free choice of circle center position allows freedom for random undersampling the k-space to accelerate acquisition by CS. This flexibility is particularly important for ${}^1$H-MRSI at 7T and beyond, due to the increased spectral bandwidth required which limits the duration of $k$-space trajectories. In addition, free choice of circle position should enable FoV with different extent along the axial dimensions for additional time saving, which cannot be achieved by concentric, rosette and spiral trajectories.

ECCENTRIC flexibility in setting FoV, resolution and spectral bandwidth by varying the circle radius and CS undersampling to optimize SNR and acquisition time is shown in Supplementary Fig. S2-S4. These results indicate that, when using the same image resolution and acquisition time, ECCENTRIC provides a higher SNR for protocols that use smaller circle radii and higher acceleration compared to protocols using larger circle radii and lower acceleration. This flexibility, allows to adapt ECCENTRIC acquisition to resolutions required across a range of FoV. Although, we demonstrated ECCENTRIC for human brain imaging, this method can be used for ultra-high resolution metabolic imaging of mice brain where submillimeter resolution is required for a centimeter FoV. 

There are some limitations in the current implementation. In particular, reconstruction time of 3D ECCENTRIC data requires several hours. For example, using a $64\times 64 \times 31$ matrix size for $3.4$~mm isotropic, the water removal step took 1 hour, the CS-SENSE-LR reconstruction took 3 hours on a GPU (or 12 hours on a CPU), and the parallel LCModel fitting took 1 hour on a high-performance server such as the Dell PowerEdge R7525 (with 64 cores of 2.9GHz and 128M cache, 512 GB RAM, and 3 NVIDIA Ampere A40 GPUs). This computation time may be considered relatively long for routine clinical applications. Also, at the moment subject motion or scanner drift is not corrected during ECCENTRIC acquisition, which may increase the variability of metabolite quantification. The metabolite concentrations were not provided in absolute units such as milimolar but expressed in institutional units (IU) relative to the water reference, which still provides comparable values across subjects and scanners. We note that for absolute quantification of FID-MRSI data only the T1 relaxation correction is needed, while ultra-short ($<$ 1ms) echo-time makes T2 relaxation negligible. Future improvements of our method will seek to speed-up the reconstruction and provide results faster to increase usability and increase robustness of data acquisition with regard to motion and hardware instability.

In summary, we have introduced ECCENTRIC an advanced acquisition-reconstruction method for MRSI that pushes the boundaries of spatial and temporal capabilities for \textit{in vivo} metabolic imaging. Although here we specifically demonstrated ECCENTRIC for MRSI at 7T ultra-high field, this method is not limited to this field and could be used at higher ($\ge$9.4T) and lower (3T) fields. ECCENTRIC has exhibited exceptional performance in metabolite mapping in healthy volunteers, showcasing high-quality results. Anticipating that ECCENTRIC will pave the way for novel advancements in neuroscience, we envision its potential to provide detailed insights into brain neurochemistry in both healthy and pathological conditions. This innovation is poised to address crucial questions and facilitate groundbreaking discoveries in both fundamental research and clinical studies.

\nopagebreak[4]

\subsection*{Funding}
This work was funded by the U.S. National Institutes of Health through National Cancer Institute (NCI/NIH) grant R01CA211080 and R01CA255479 (O.C.A), Swiss National Science Foundation grant IZSEZ0\_188859 (A.K.), and by Austrian Science Fund (FWF): KLI 1106, P 34198 and J 4124 (B.S, L.H. and W.B.). This research used the imaging equipment at Athinoula A. Martinos Center for Biomedical Imaging provided by the Center for Functional Neuroimaging Technologies, P41EB015896, a P41 Biotechnology Resource Grant supported by the National Institute of Biomedical Imaging and Bioengineering (NIBIB), National Institutes of Health. 

 \setcounter{table}{0}
\renewcommand{\thetable}{S\arabic{table}}%
 \setcounter{figure}{0}
 \renewcommand{\thefigure}{S\arabic{figure}}%
 
\pagebreak
\section{Supplementary Material :  Additional data} 

\section{Additional data} 

\subsection{ECCENTRIC water imaging in healthy volunteers}
\begin{figure}[H]
\centering
\includegraphics[width=1\textwidth]{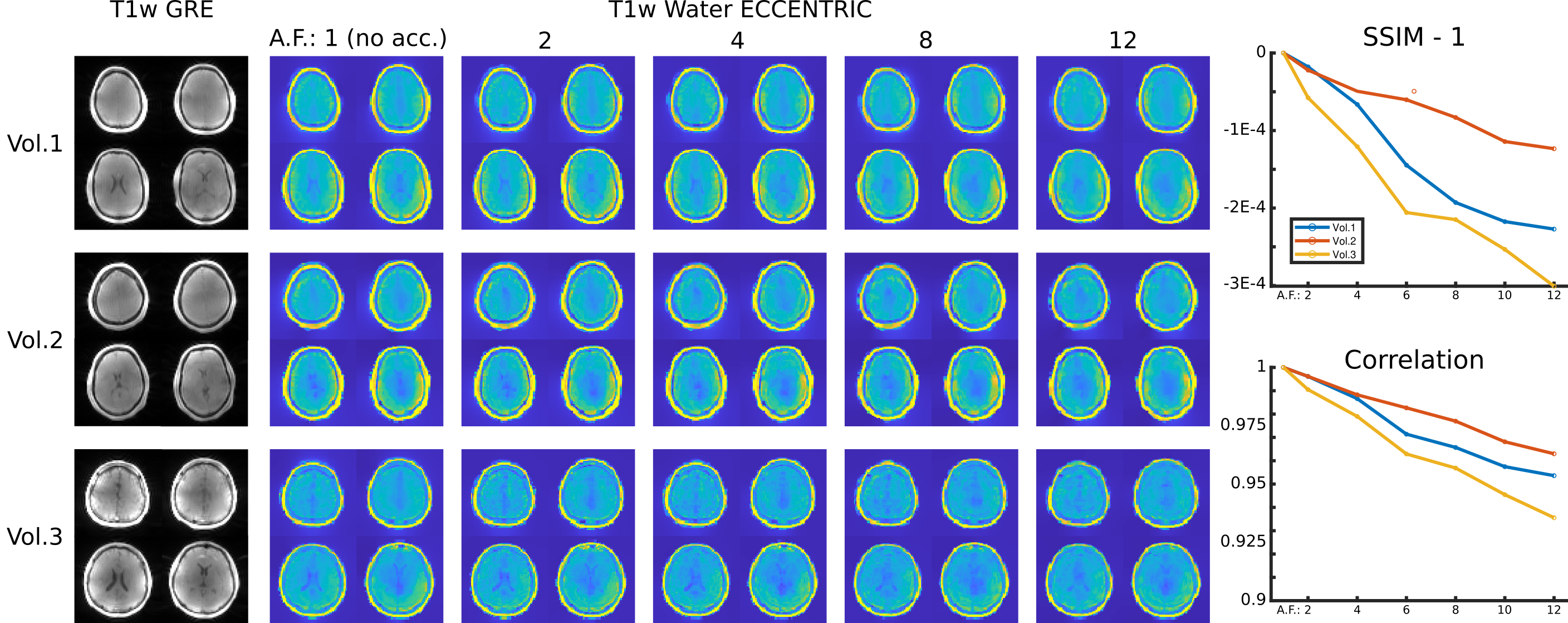}
\caption{Water imaging of human brain in healthy volunteers using 3D-ECCENTRIC with $3.4$mm isotropic voxel size. Left, images obtained by $T_1$-weighted 3D GRE acquired with matched FA, TR, TE and spatial resolution. Middle, water images obtained with 3D-ECCENTRIC for acceleration factors 1-12. Right, SSIM and correlation factors for accelerated ECCENTRIC water images are calculated considering the fully sampled image (AF=1) as ground truth. Four slices are shown for each volunteer.}
\label{fig:WaterVols}
\end{figure}

To investigate the ability of ECCENTRIC to image brain structure, we performed water imaging in several healthy volunteers using 3D FID-MRSI ECCENTRIC. The water suppression was turned off, and we used the 3.4~mm protocol with larger FA = {40\textdegree} and shorter TR = 100 ms to produce $T_{1}$ weighed images. ECCENTRIC k-space data were acquired fully sampled (AF=1) and the acceleration was obtained by retrospective CS undersampling in post-processing. Results in Fig. \ref{fig:WaterVols} show that fully sampled ECCENTRIC images reveal similar brain structure as $T_1$-weighted GRE images acquired with matched spatial resolution and tissue contrast. Considering the fully sampled (AF=1) ECCENTRIC as ground truth, the SSIM $\ge$ 0.99 and correlation factor $\ge$ 0.92 for images obtained with AF = 1-12. Visually, almost no difference can be observed between images obtained with AF = 1-4. For AF $\geq$ 8 the noise level increases, which interferes with fine structural details.

\section{Optimizing ECCENTRIC for high SNR and accelerated high-resolution MRSI} 

By design, the k-space acquisition by ECCENTRIC is characterized by: the circle radius (CR), compress sensing acceleration factor (AF), image matrix size (MS) and field-of-view (FoV). In addition, acquisition of the time dimension (FID) for spectroscopy is characterized by the spectral bandwidth (SW), the dwell-time, and the number of time points. In particular, for spectral-spatial encoding (SSE) there is a dependency between the spectral bandwidth, field-of-view and image resolution. Compared to other SSE schemes, ECCENTRIC allows very high flexibility in the choice of SW, FoV and MS, which is particularly needed at ultra-high field (7T and beyond) and to operate within the technical limits of the gradient system minimizing electrical, mechanical and thermal stress. 

Importantly, the ECCENTRIC flexibility can be used to optimize the SNR and acquisition time while pushing the image resolution. ECCENTRIC parameters have different impact on the measured signal-to-noise ratio (SNR) and acquisition time (TA) as highlighted in the following table:

\begin{table}[H]
\centering
    \begin{center}
    \begin{tabular}{| c | c | c|} 
        \hline
            Change  \textbackslash \: Effect 	&  SNR	& TA \\ \hline
            Circle Radius $ \nearrow $ & $ \searrow $ &	$ \searrow $  \\ 
CS Acceleration $ \nearrow$	& $\rightarrow$  (\text{image smoothness} $ \nearrow $)	 & $\searrow $  \\ 
Matrix Size $\nearrow $	&  $\searrow$ & $\nearrow $  \\ \hline
    
        \hline
    \end{tabular} 
    \caption{  }
    \label{table:CRLB_SNR_FWHM}
    \end{center}
\end{table}

3D ECCENTRIC MRSI can be optimized by reducing CR and thus increasing the sampling density in k-space, which can be designed to sample more the center of k-space to increase SNR. In addition, the CR reduction allows a large range of spectral windows by controlling the gradient slew rate as needed based on the image resolution. However, the reduction in CR requires a higher CS acceleration for an equivalent acquisition time. As we showed, CS acceleration up to 4 provides high quality metabolite images, and this can be traded to optimize SNR with smaller CR. 

\begin{figure}[H]
\centering
\includegraphics[width=1\textwidth]{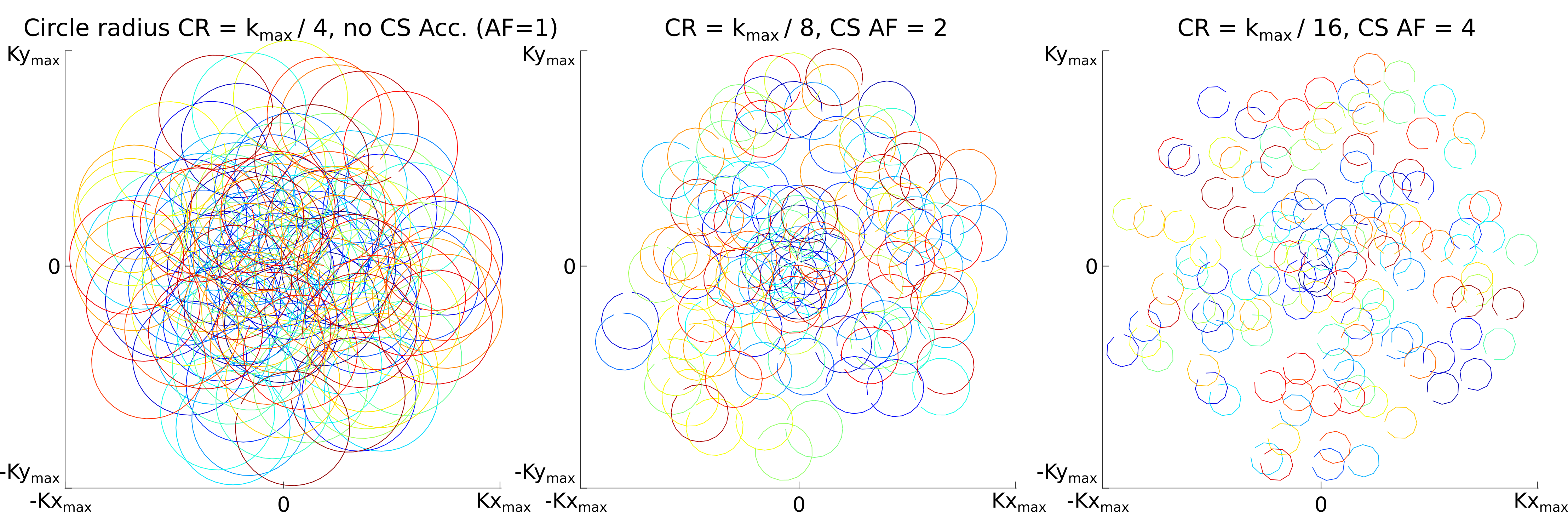}
\caption{ECCENTRIC k-space sampling for various circle radii (CR) and compress sensing acceleration (AF). }
\label{SDfig:TrajDiffCirc}
\end{figure}

To explore the flexibility of ECCENTRIC parameters for SNR optimization, three acquisitions with the same isotropic resolution (5mm) and acquisition time (14min) were acquired with different ECCENTRIC circle radii (CR) and CS acceleration factors: 1) CR = kmax/4 and AF = 1; 2) CR = kmax/8 and AF = 2; 3) CR = kmax/16 and AF = 4. Sampling patterns of the k-space are shown in Fig. \ref{SDfig:TrajDiffCirc}.

\begin{figure}[H]
\centering
\includegraphics[width=0.9\textwidth]{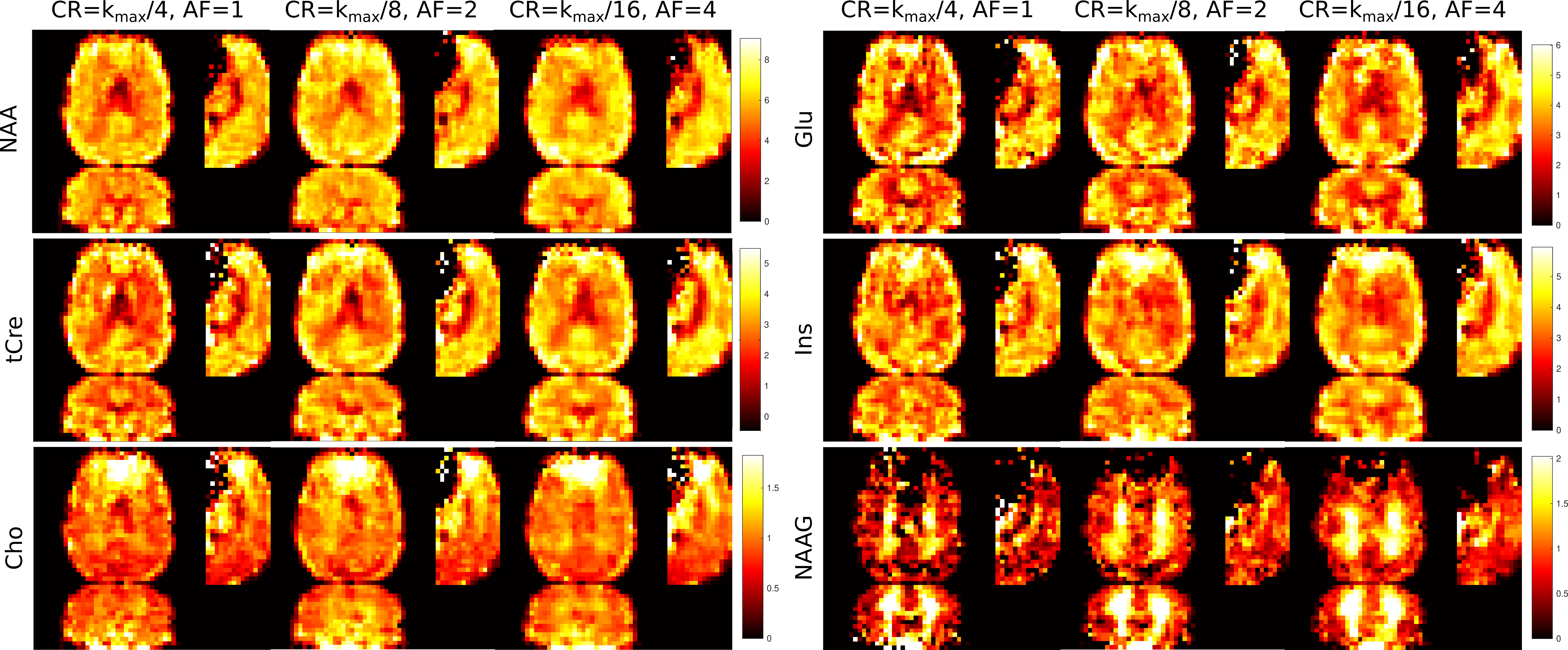}
\caption{Metabolite maps of NAA, total Creatine, total Choline, Glu, Ins and NAAG produced from 3D ${}^1$H-FID-MRSI ECCENTRIC acquisitions at 5~mm isotropic image resolution in 14 min with various circle radii CR and compress sensing AF. }
\label{SDfig:MetabMaps}
\end{figure}

Metabolite maps obtained in a healthy volunteer are presented in Fig. \ref{SDfig:MetabMaps}. Higher SNR can be noticed for the six metabolites as the circle radii is decreased, while only minor blurring is apparent at the highest acceleration. 

\begin{figure}[H]
\centering
\includegraphics[width=1\textwidth]{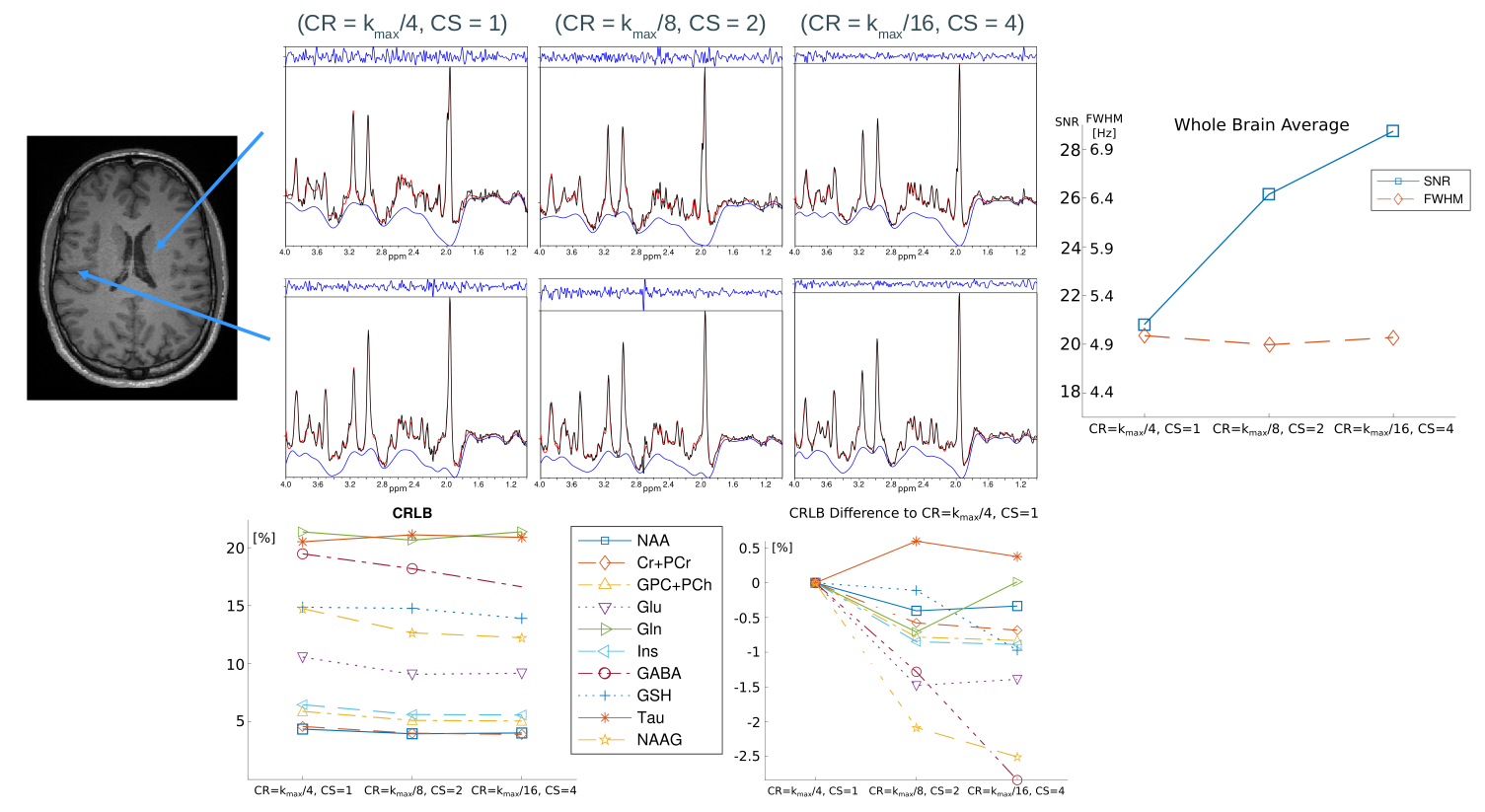}
\caption{Quantitative analysis of 3D ${}^1$H-FID-MRSI ECCENTRIC acquisitions with various circle radius CR and compress sensing AF.  }
\label{SDfig:Spectra}
\end{figure}

Quantitative analysis in Fig. \ref{SDfig:Spectra} shows that decreasing CR results in a notable gain in metabolite SNR of +~30\% for kmax/8 and +~40\% for kmax/16 relative to kmax/4, respectively. In the same time, the linewidth is stable across the different protocols. The increase in SNR enables more precise metabolite quantification resulting in lower CRLB for spectral fitting, especially for the low signal metabolites. Among the 3 protocols, the protocol with CR = kmax/8 and AF = 2 showed the best performance with a marked increase in SNR and little visible blurring on metabolite maps. These results demonstrate that CR and AF allow SNR optimization of 3D ECCENTRIC MRSI for a desired image resolution and acquisition time.

\bibliography{Biblio}

\end{document}